\documentclass[a4paper]{emulateapj}
\usepackage{amstext}
\usepackage{amsmath}
\usepackage{amssymb}
\usepackage[normalem]{ulem}
\usepackage[colorlinks=true,linkcolor=blue,citecolor=blue]{hyperref}
\begin{document}



\title{The dynamics, appearance  and demographics of relativistic jets triggered by  tidal disruption of stars in quiescent supermassive black holes}

\author{Fabio De Colle, James Guillochon,  Jill Naiman and Enrico Ramirez-Ruiz\altaffilmark{1}}
\altaffiltext{1}{TASC, Department of Astronomy and
  Astrophysics, University of California, Santa Cruz, CA
  95064, USA}
\email{(fabio, jfg, jnaiman, enrico)@ucolick.org}


\begin{abstract}
We examine the consequences of a  model in which relativistic jets can be triggered in 
quiescent massive black holes  when a geometrically thick and hot accretion disk forms 
as a result of the tidal disruption of a star. 
To  estimate the power, thrust and lifetime of the jet, we use  the mass accretion history onto the black hole as calculated by detailed
hydrodynamic simulations of the tidal disruption of stars. We go on to determine the states of the interstellar medium in various types of quiescent galactic nuclei, and describe how this external matter can affect jets propagating through it. We use this information, together with a two-dimensional hydrodynamic model of the structure of the relativistic flow, to study  the dynamics of the jet, the propagation of which  is regulated by the density stratification of the environment and by its  injection history. The breaking of symmetry involved in transitioning from one to two dimensions is crucial and leads to qualitatively new phenomena.
At early times, as the jet power increases,  the high pressure of the cocoon collimates the jet, increasing its shock velocity as compared to that of spherical  models.  
We show that small velocity gradients, induced near or at the source, steepen into internal shocks and provide  a
source of free energy  for particle acceleration  and  radiation  along the jet's channel.
The jets terminate at a working surface where they interact strongly with the surrounding
medium through a combination of shock waves and instabilities; a continuous flow of relativistic fluid emanating from the nucleus supplies this region with mass, momentum and energy.  Information about  the $t^{-5/3}$ decrease in power supply propagates within the jet at the internal sound speed. As a result, the internal energy  at the jet head continues to accumulate until long  after the peak feeding rate is reached. An appreciable  time delay is thus expected  between peaks in the short-wavelength radiation emanating  near the jet's origin and the long-wavelength emission produced at the head of the jet.  Many of the observed properties of  the Swift 1644+57/GRB 110328A event can be understood as resulting from accretion onto and jets driven by a $10^6M_\odot$ central mass black hole following the disruption of sun-like star.
With the inclusion of a stochastic contribution to the luminosity due to variations in the feeding rate driven by instabilities near the tidal radius, we find that our model can explain the X-ray light curve  without invoking a rarely-occurring deep encounter. In conjunction with the number density of black holes in the local universe, we hypothesize that the conditions required to produce the Swift event are not anomalous, but are in fact representative of the jet-driven flare population arising  from tidal disruptions.
\end{abstract}



  \section{Introduction}
  \label{sec:introduction}

Relativistic jets accelerated from  compact objects, such as neutron stars or  black holes, 
are suspected to produce many of the observational signatures associated with high energy phenomena.  
Objects thought to produce them include radio galaxies and quasars \citep{begelman84}, 
microquasars \citep{mirabel99} and gamma ray bursts \citep{gehrels09}.
An important difference between jets of gamma ray bursts  and the better studied radio jets 
of quasars or microquasars is that active quasars inject energy over extended periods 
of time into the jet while gamma ray burst sources are impulsive. 
Therefore, quasar jets remain highly collimated throughout their lifetimes, while gamma ray burst 
jets decelerate and expand significantly once they become non relativistic \citep[][]{ayal01, ramirez-ruiz10}.

\begin{figure*}
\centering
\includegraphics[width=\linewidth]{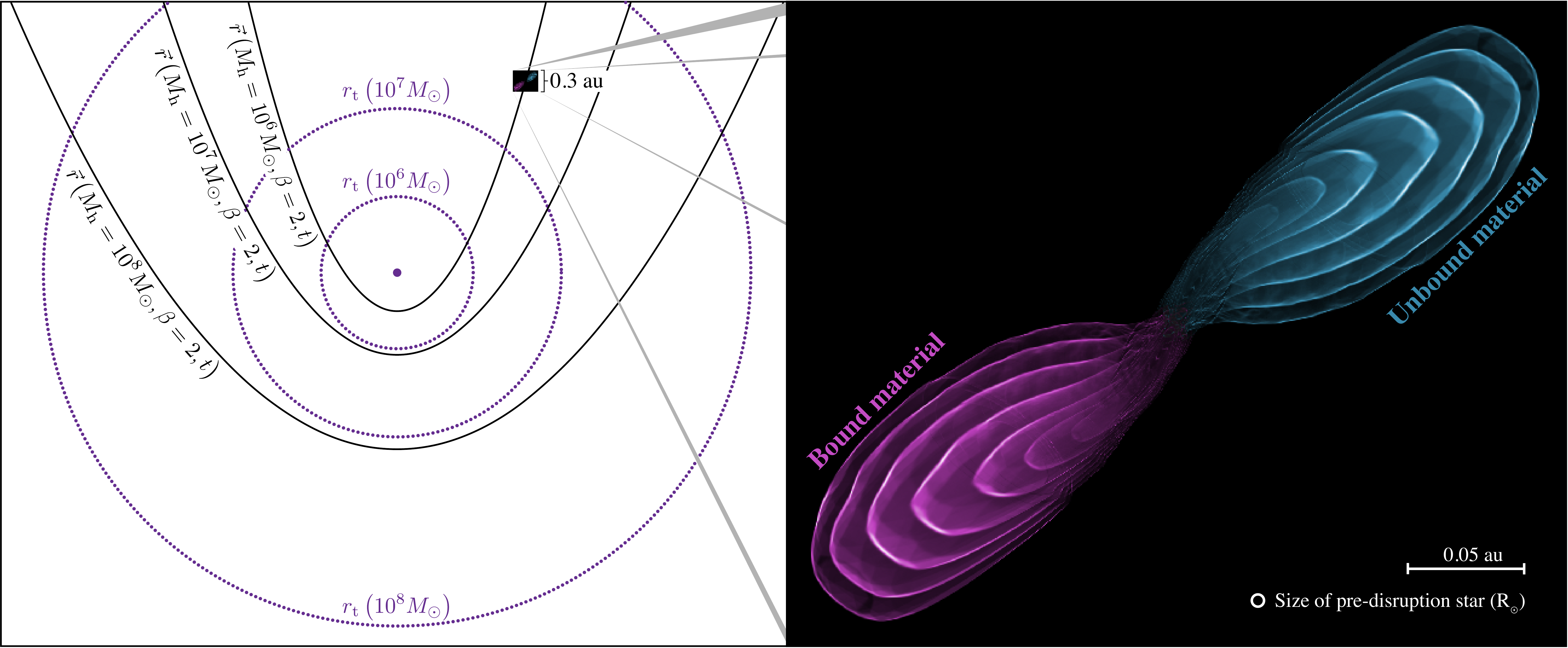}
\caption{Diagram of the  complete tidal  disruption of a Sun-like star for several black 
hole masses. The left panel 
shows the system geometry for three black hole masses, $M_{\rm h} = \left(10^{6}, 10^{7}, 
10^{8} M_{\odot}\right)$. Shown in black are the Newtonian orbital trajectories 
$\vec{r} \left(M_{\rm h}, \beta, t\right)$ for a $\beta = 2$ encounter, which we find to be 
fully disruptive in our hydrodynamical simulations. The tidal radius $r_{\rm t}$ for each 
$M_{\rm h}$ is shown by the dotted lines. Inset into the diagram along each orbital 
trajectory is a to-scale rendering of the orbital debris 4 hours after pericenter, generated 
from a single simulation with $M_{\rm h} = 10^{6} M_{\odot}$. The right panel shows a zoom-in 
of this debris, with contours showing constants values of $\rho e_{\rm h}$, where $e_{h}$ is 
the specific binding energy to the black hole, with the contours ranging from $10^{12}$ ergs 
cm$^{-3}$ to $10^{16}$ ergs cm$^{-3}$ with logarithmic spacing. Material that is bound to the
 black hole after the encounter is shown in magenta, whereas unbound material is shown in cyan. 
The symmetry of the debris demonstrates the nearly symmetrical tide imposed by the black hole
 at closest approach.} 
\label{fig:tidaldis}
\end{figure*}

Recent observations of the Swift 1644+57/GRB 110328A event \citep{bloom11, burrows11, levan11}
and of the Swift J2058.4+0516 event \citep{cenko11} have been interpreted 
\citep{bloom11, giannios11, metzger11, miller11, shao11, zauderer11, vanVelzen11,cannizzo11} 
as evidence that relativistic jets can be triggered in quiescent massive black holes when an orbiting  Sun-like star, 
owing to the cumulative effect of encounters with other stars \citep{frank77}, gets too close to the 
black hole and is tidally disrupted\footnote{Alternative models have also been 
considered \citep{socrates11,krolik11,quataert11,ouyed11, woosley11}}.

Because the duration of the event is determined by the timescale at which the 
most bound material returns to pericenter  forming an accretion disk 
\citep{evans89, rees98, rosswog09,ramirez-ruiz09,lodato09,guillochon09}, 
flaring black hole candidates in nearby galaxies offer the possibility of watching 
the evolution of a quasar-like object through many stages of its life in a 
time span of few months  or years \citep[e.g.,][]{berger11} 
rather than waiting the millions of years necessary to observe changes in extragalactic 
objects.

Much of our effort in this paper is therefore dedicated to study the dynamics 
of jets triggered by tidal disruption. Some of the questions at the forefront of 
our attention include the effects of the external medium and the degree to which the 
jet dynamics are modified by their energy injection histories. Because the mass accretion rate  onto a black hole that is fed by  
tidal disruption is far from being steady and there is not a simple 
prescription for the surrounding density  stratification, self-similar solutions fail to provide 
an accurate description of the jet dynamics, and thus    
simulations are required.
The mass accretion history onto the black hole as well 
as the jet lifetime  are calculated in Section \ref{sec:rates} using 
detailed hydrodynamical simulations of  the tidal disruption of Sun-like stars. 
The character of the external medium 
responsible for  shaping the evolution and morphology of the jets is reviewed 
in Section \ref{sec:medium}.
Detailed hydrodynamic calculations of the evolution of jets triggered by tidal  
disruption are presented in Section \ref{sec:jet}, together with a brief description 
of the numerical methods and the initial model.  
Finally, a tidal disruption model for  the Swift 1644+57/GRB 110328A event  is presented in Section \ref{sec:discussion}, followed by a discussion on how the discovery of flaring candidates in nearby galaxies 
by {\it Swift} will help elucidate the demography of the dormant black hole 
population.

  \section{Feeding rates and jet activity following tidal disruption}
  \label{sec:rates}

  \subsection{Tidal Disruption Simulations}
Our formalism for calculating the rate of mass return to the black hole $\dot{M}_{\rm h}$ 
after the disruption is identical to the method presented in \citet{guillochon11}, except 
for the initial conditions where we take the star to be a 1 $M_{\odot}$ star described by 
a polytropic index $n = 3$ and an adiabatic equation of state with $\gamma = 5/3$. Our 
disruption simulations are performed using the FLASH hydrodynamics framework \citep{fryxell00}, 
which includes an adaptive mesh refinement scheme that permits the wide range of scales 
necessary to resolve the star and the debris streams simultaneously. During the disruption, 
the debris tails are adaptively refined based on their density relative to the maximum 
density in the simulation $\rho_{\max}$ at each time step, with a cutoff
density for the lowest refinement level of $10^{-19}$ times the maximum density. All matter 
with $\rho > 10^{-3} \rho_{\max}$ is refined to the highest level. The disruption is 
performed in a box that is several orders of magnitude larger than the initial star's 
radius of $r_{\ast} = R_{\odot}$, which is done to facilitate tracking of the tidal debris 
over long enough timescales for hydrodynamical effects to no longer play a major role
in determining the distribution of mass as a function of binding energy.

  \subsection{Deriving $\dot{M}_{\rm} (M_{\rm h})$}

\begin{figure}
\centering
\includegraphics[width=\linewidth]{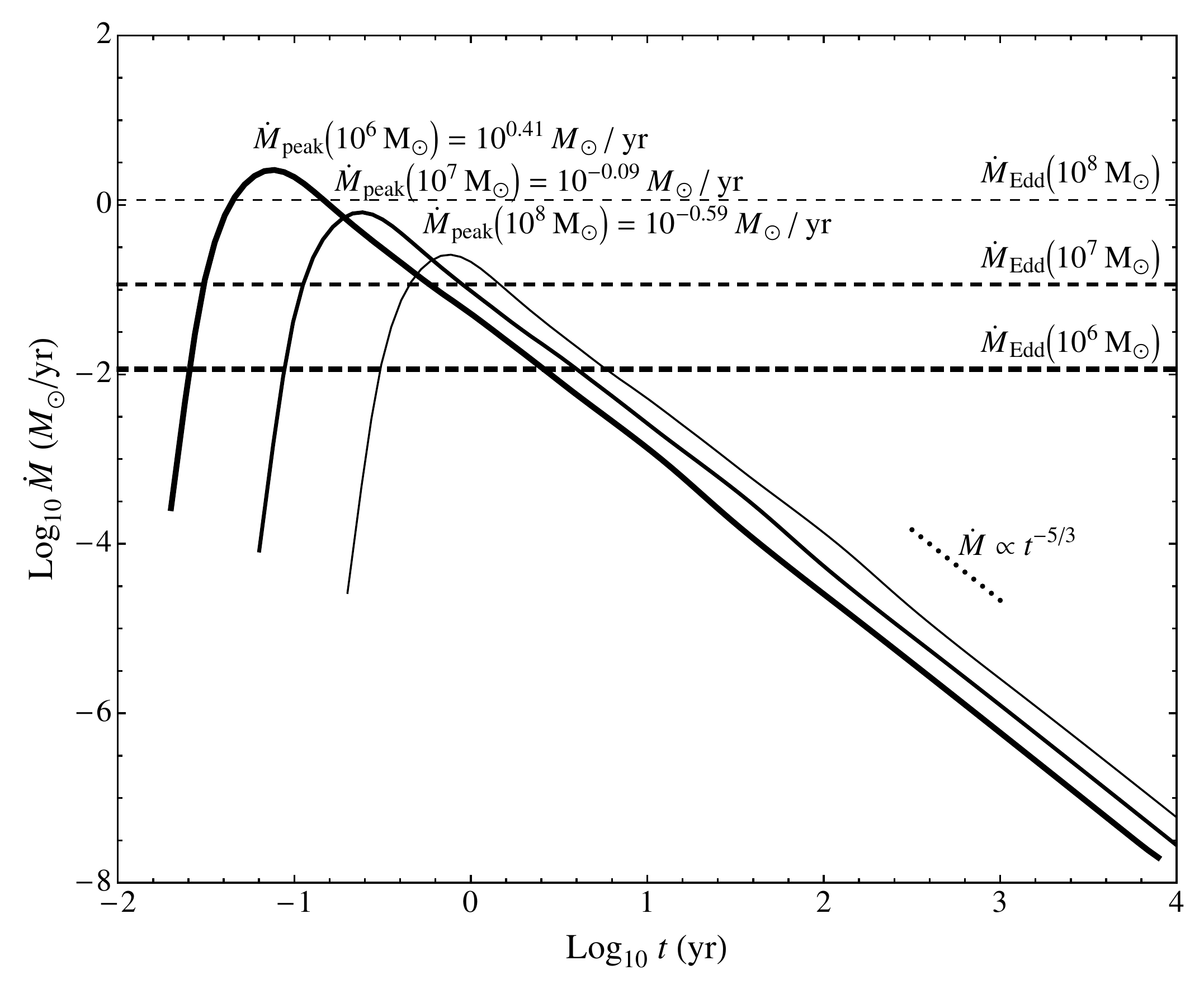}
\caption{Rate of return of mass $\dot{M}$ to the black hole following a $\beta = 2$ encounter, 
displayed as solid lines of decreasing thickness 
for increasing $M_{\rm h}$. As the time of the 
encounter is independent of $M_{\rm h}$ at constant $\beta$, and the forces experienced along 
the trajectory are nearly identical, the rate of mass return simply scales with the spread of 
binding energy across the debris, \(\dot{M} \propto M_{\rm h}^{-1/2}\). The feeding rate 
rapidly converges to the predicted $t^{-5/3}$ power law, as shown by the short dotted line 
segment. Dashed lines show the Eddington feeding limit $\dot{M}_{\rm Edd}$ as a function of 
$M_{\rm h}$, assuming an  efficiency $\epsilon = 0.2$. The larger the mass of the black hole, 
the shorter the amount of time the black hole remains above a given $\dot{M}$. A disruption 
of a $1 M_{\odot}$ star by a $10^{8} M_{\odot}$ black hole never exceeds the Eddington limit.}
\label{fig:dmdedens}
\end{figure}

To calculate the rate of mass return to the black hole $\dot{M}_{\rm h}$ 
as a function of the black hole mass $M_{\rm h}$, we performed a single simulation 
of the disruption of a 1 $M_{\odot}$ star by a $10^{6} M_{\odot}$ black hole with impact 
parameter $\beta \equiv r_{\rm t}/r_{\rm p} = 2$, where $r_{\rm t}$ and $r_{\rm p}$ are the 
tidal and the periastron radius, respectively (see Figure \ref{fig:tidaldis}). 
This encounter is deep enough to leave no surviving core \citep{guillochon12}. 
To accurately model the disruption, we perform the simulation using a cubical volume with 
a basic grid of $8^3$ cells (along each Cartesian coordinate axis), with a width of 
$4\times 10^{14}$ cm, and 16 levels of refinement, resulting in the initial diameter ($= 2 R_\sun$)
of the star being resolved by 90 grid cells. The shape of $\dot{M}_{\rm h}$ in the Newtonian 
approximation for full disruptions is only affected by the degree of symmetry in the tides 
\citep{guillochon11} and the star's initial density profile \citep{lodato09,ramirez-ruiz09}. As shown in 
\citet{guillochon11}, tides are very symmetrical for black holes with $M_{\rm h} \gtrsim 10^{6} M_{\odot}$, 
with the maximum force difference between the near- and far-side of the star being 6\% for a $\beta = 2$ encounter involving a $M_{\rm h} = 10^{6} M_{\odot}$, 
and even less for black holes of larger mass. Assuming that the shape of $\dot{M}_{\rm h}$ 
does not vary for black holes of mass $\ge 10^{6} M_{\odot}$, $\dot{M}_{\rm h}\left(M_{\rm h}\right)$ 
can be related to our benchmark simulation via
\begin{align}
\dot{M}_{\rm h}\left(M_{\rm h}, t\right) &= \dot{M}_{\rm h}\left(10^{6} M_{\odot}, t^{\prime}\right)\left(\frac{M_{\rm h}}{10^{6}M_{\odot}}\right)^{-1/2}\\
t^{\prime} &= t\left(\frac{M_{\rm h}}{10^{6}M_{\odot}}\right)^{-1/2},
\end{align}
where $t$ and $t^{\prime}$ are the times since disruption for a black hole of mass $M_{\rm h}$ 
and our benchmark simulation, respectively \citep{evans89}. 
Our simulation assumes that the typical disrupted 
star has a structure similar to that of the Sun, but stars of different masses and ages can 
have different degrees of central concentrations \citep{tout96}. \citet{lodato09} 
showed that for full disruptions the star's initial density profile primarily 
affects $\dot{M}_{\rm h}$ at early times, but 
that all reasonable initial stellar profiles eventually lead to a power-law decay where 
$\dot{M}_{\rm h} \propto t^{-5/3}$ at late times. Therefore, we assume that a typical disruption of a 1 
$M_{\odot}$ by a black hole of mass $M_{\rm h} \ge 10^{6} M_{\odot}$ can be reasonably modeled
 by a single simulation using the above scalings, which yields the $\dot{M}_{\rm h}$ shown in 
Figure \ref{fig:dmdedens}.

\begin{figure*}
\centering
\includegraphics[width=\linewidth]{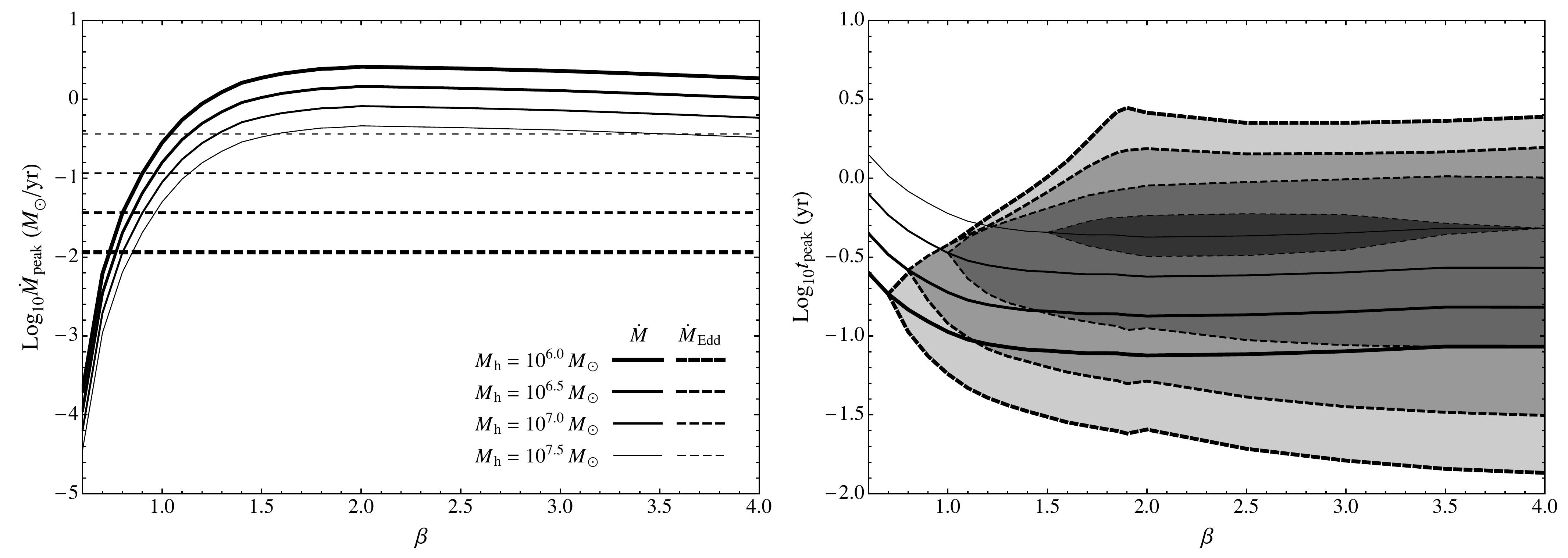}
\caption{Peak accretion rate and time at which accretion peaks as a function of $\beta$ for simulations of a $1 M_{\odot}$ stellar disruption. The solid lines in the left panel show the peak accretion rate $\dot{M}_{\rm peak}$ as a function of $\beta$ for several $M_{\rm h}$, with the largest accretion rates being produced for $\beta = 2$ encounters. The corresponding Eddington accretion rates $\dot{M}_{\rm Edd}$ are shown as dashed lines. The right panel shows the time $t_{\rm peak}$ at which this peak accretion rate is achieved (solid lines of varying thickness), where $t$ is measured relative to the time of pericenter. The span of time for which the black hole exceeds $M_{\rm Edd}$ is shown by the gray contours which are bounded by the dashed curves. Disruptions occurring around black holes of larger mass peak at later times relative to pericenter, and exceed the Eddington threshold for a much shorter duration than their low-mass counterparts.}
\label{fig:mdottime}
\end{figure*} 

  \subsection{The Expected Jet Lifetime}

The returning gas does not immediately produce a flare of activity from the black hole. First
material must enter quasi-circular orbits  and form an accretion torus \citep{ramirez-ruiz09}. Only then
will viscous effects release enough binding energy to power a flare. Once the torus is formed, it will evolve under
the influence of viscosity and radiative cooling, although the viscosity
 would have to be implausibly low (i.e. the usual viscosity
 dissipation time $t_\nu$ for a thick disk would be $\alpha^{-1}$) for the bulk of the mass to be stored for longer than
 $t_{\rm peak}$ in a reservoir at $r \approx r_{\rm t}$.  After $t_\nu \ll t_{\rm peak}$ the mass accretion rate  would
 continue to fade as $\dot{M}_{\rm h}$. 

Little is known about the relation between
jet production in  supermassive accreting black holes  and the state of their constituent accretion disks. 
However, an  association between hot thick accretion flow with the strongest jets, as observed in binary black holes,  is expected \citep[e.g.][]{krolik2012}. 
It is the poloidal field protruding from the disk that is thought to drive the jet and because its strength increases with disk thickness, jets are expected to be  stronger in thick disks than in thin ones \citep{meier01}.
In GX 339-4, for example, a jet is produced when the X-ray source is in the low/hard state. In this state, the disk temperature is $T\sim 10^9$K,  suggesting a thick disk. 
On the other hand, when the source enters the high/soft state (with  a thin $10^7$ K disk)
 the jet radio emission disappears to a level at least  tens of times weaker \citep{fender99}. 
In what follows,  we make the assumption that jets will be produced only when the accretion disk is geometrically thick and hot. 

It is clear from Figure~\ref{fig:mdottime}  that most of the debris
would be fed to the black hole far more rapidly than it could be accepted if the radiative efficiency
was high. As the disk material advects onto black hole, we thus assume a relativistic jet 
with $L_{\rm j} (t)\propto  \dot{M}_{\rm h} (t)$ can be powered
for as long as  $ \dot{M}_{\rm h} (t)\gtrsim \dot{M}_{\rm Edd}$. 

\begin{figure}
\centering
\includegraphics[width=\linewidth]{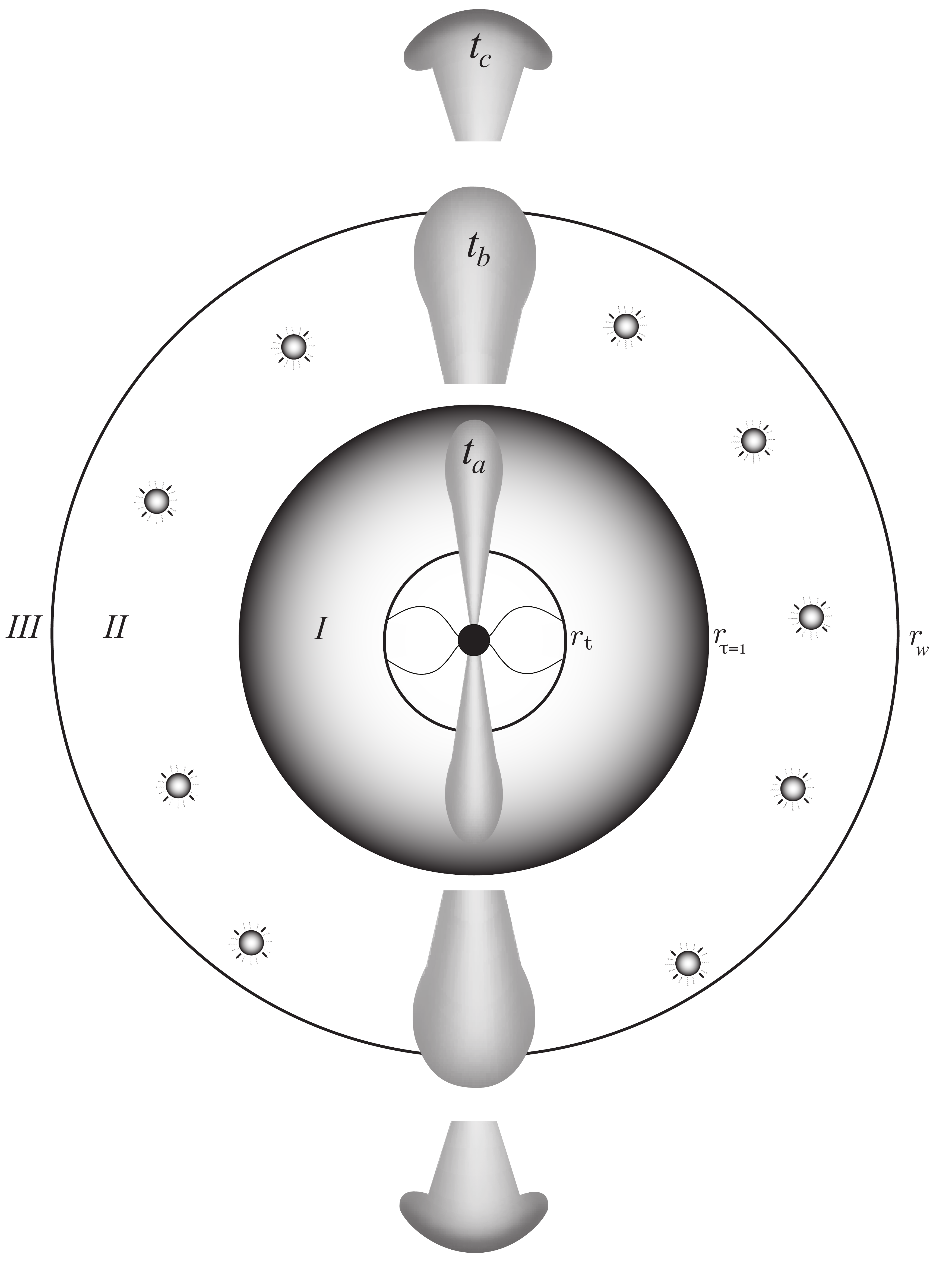}
\caption{Schematic diagram of a relativistic jet propagating through the nucleus of a quiescent galaxy, illustrating the nomenclature of
the external medium inferred to exist  and   the expected  dynamical stages.
Initially, a nearly hydrostatic envelope around the black hole,  within which  radiation pressure is dominant ($r\leq r_\tau$),  will  impede the advancement of the jet (region $I$). After  emerging from the optically thick envelope, the jet is believed to move into a medium whose properties are primarily determined by the emanating stellar outflows  within  the nuclear cluster (region $II$).   A jet emerging from the dense stellar core  may pass through a region of steadily decreasing ambient pressure for up to several tens of parsecs (region $III$).  Jet material travels along a channel of its own making. The speed with which the head of the channel advances into the surrounding medium  is obtained by balancing the momentum flux in the jet against the momentum flux of the surrounding medium. A continuous flow of relativistic fluid supplies this region with mass, momentum, energy, and magnetic flux 
 for as long as $t_{\rm a} \lesssim t_{\rm Edd}$. Information about a sizable decrease  in power supply propagates into the jet at the internal sound speed (for $t \approx t_{\rm b}$), reaching the head when the jet has traveled a further distance $r(t_{\rm c})$. Once these transient phase terminates, the bulk of the jet material expands freely.
    }
\label{fig:diagram}
\end{figure} 

  \section{Properties of the Surrounding Medium}
  \label{sec:medium}

With the exception of our own galactic center and M31 where there are observational constraints on the extent of  the confining  gas  on scales $r \leq 10$ pc within the nucleus \citep{quataert04, garcia10}, little is known about the character of the medium  surrounding  black holes  in the cores of inactive galaxies.  We can thus place only  somewhat model-dependent limits on the surrounding density and pressure structures. The jet's advancement would be initially impeded near $r_{\rm t}$, where  the returning  bound material  forms a nearly hydrostatic envelope around the black hole \citep{loeb97}, within which  radiation pressure is dominant (region $I$ in Figure~\ref{fig:diagram}). After  the jet has broken free from the optically thick envelope, it  passes through an extended  region of steadily decreasing ambient pressure, whose properties are primarily  shaped by stellar  wind  collisions within the dense core (region $II$ in Figure~\ref{fig:diagram}).  The most rapid drop in pressure probably occurs outside the dense core (region $III$ in Figure~\ref{fig:diagram}), where mass whose gravitational field confines the ambient gas is likely to have an extended distribution, but with density rapidly decreasing outwards. An understanding of the structure and evolution of jets triggered by tidal disruption can come only through knowledge of the properties of the material through which they propagate.  For this reason, we now consider the surrounding density and pressure profiles  in more detail.

  \subsection{Optically Thick Envelope}
  \label{sec:env}

\begin{figure*}
\centering
\includegraphics[width=\linewidth]{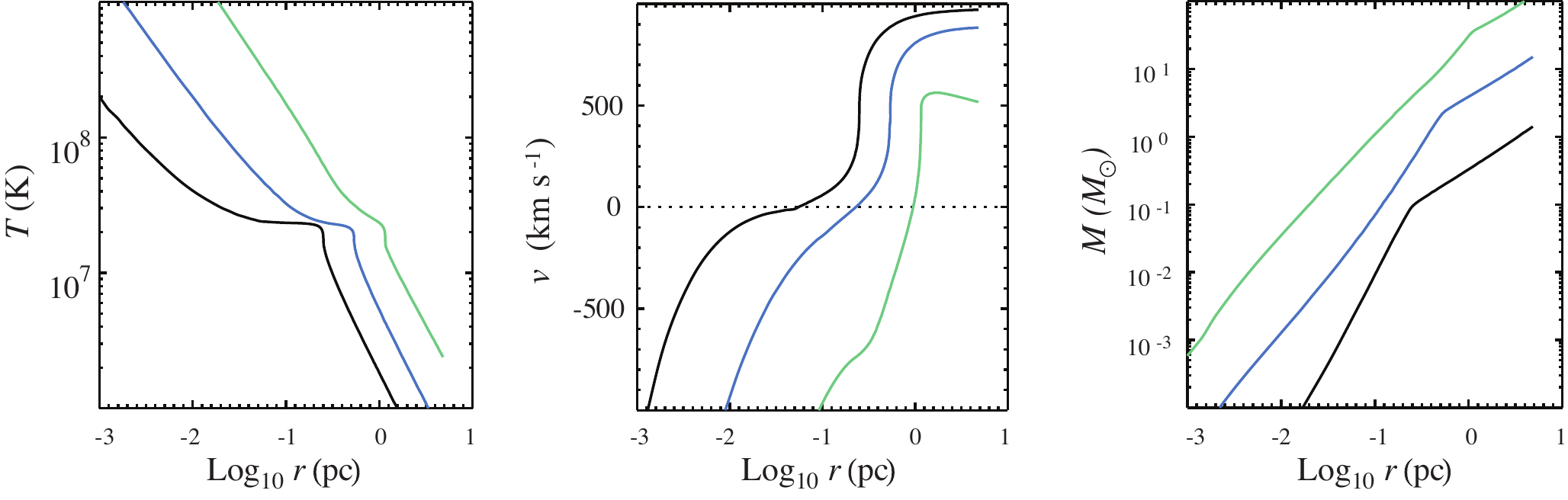}
\caption{
Comparison of the gas properties surrounding quiescent  black holes of $10^6$ (black curves), $10^7$ (blue, dashed curves) and $10^8 M_\odot$ (green, dotted curves) for $\eta=0$.  The profiles for $10^6 M_\odot$ have been derived using the stellar wind  properties observed in our own galactic nuclei, for which  $\Sigma \dot{M}_{\rm w} \approx 10^{-3}M_\odot\;{\rm yr^{-1}}$ and $v_{\rm w}\approx 10^3\;{\rm km\;s^{-1}}$ \citep{quataert04}.  In all cases, the flow at large radii (after a few sound crossing times) resembles that of a wind while interior to the
mass injection region, the gas is captured and accreted onto the black hole.  Fluid variables plotted from left to
right are the temperature, $T(r)$, the flow velocity, $v(r)$, and the integrated mass profile, $M(<r) = \int_0^r 4 \pi \rho(r') r'^2 dr'$.
 }
\label{fig:mhden}
\end{figure*} 

\begin{figure}
\centering
\includegraphics[width=\linewidth]{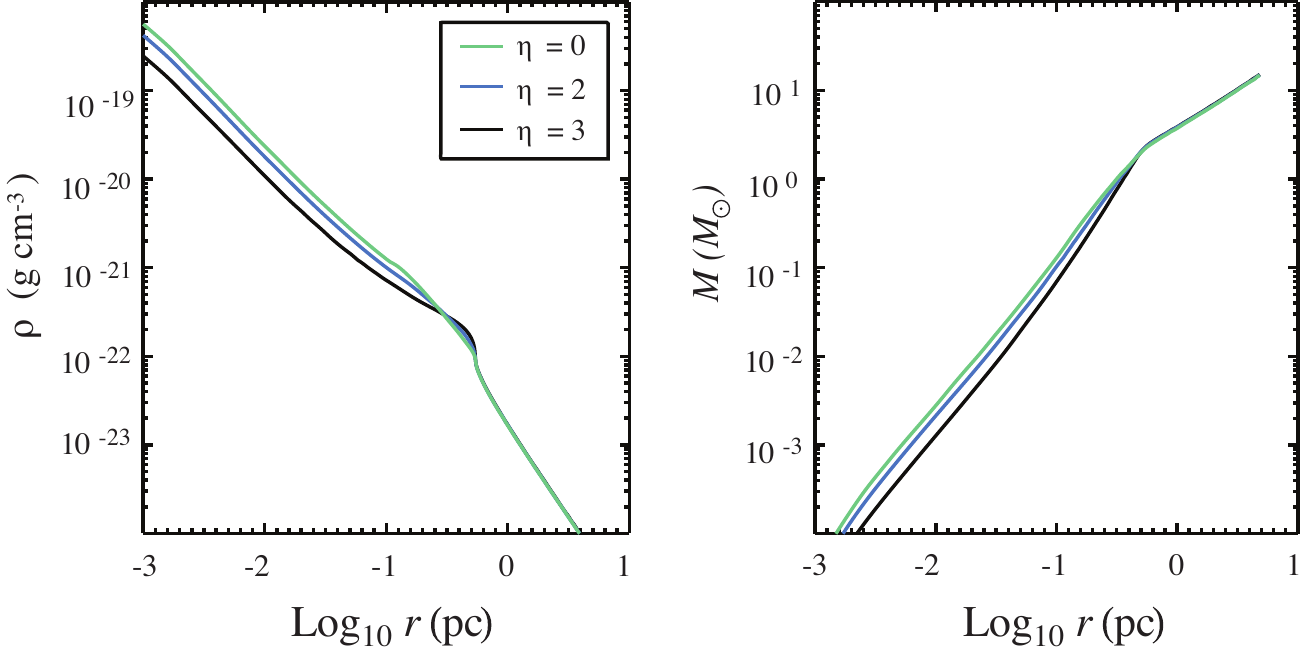}
\caption{Densities and integrated mass profiles expected around a $10^7 M_\odot$  mass black hole for different mass injection distributions, parametrized here by  $q(r) \propto r^{-\eta}$ for  $r \, \epsilon \, [r_{\rm m}, r_{\rm M}]$ . }
\label{fig:mhdeneta}
\end{figure} 

The structure of the optically thick, high entropy envelope formed as a 
result of the tidal disruption of a star by a massive black hole has been
described by \citet{loeb97}. Their results, used as initial conditions  
in Section \ref{sec:jet}, are briefly described here.
During the tidal disruption process,  about half of the stellar material 
escapes on hyperbolic orbits with speeds $\sim 9500 (M_{\rm h}/10^6M_\odot)^{1/6}$ km/s, while the rest falls back onto the black hole.  The bound gas, after pericenter passage, is on orbits which collide with the infalling stream near the original orbital
plane at apocenter, giving rise to a shock which redistributes angular momentum \citep{rosswog09,ramirez-ruiz09}. The debris raining down would, after little more than its free-fall time, settle into a disk surrounded by a radiation-dominated envelope, whose inner radius is
\begin{equation}
 r_{\rm t} = r_{\ast} \left( \frac{M_{\rm h}}{M_{\ast}} \right)^{1/3} \approx 2.15
  \times 10^{13} \left( \frac{M_{\rm h}}{10^6 M_{\odot}} \right)^{1/3}
  \mathrm{cm}
  \label{eq:rin}
\end{equation}
where $r_{\ast} \approx R_{\odot}$ and $M_\ast \approx M_\odot$. 

The structure of such an envelope is simplified by the fact that a fully ionized gas dominated by Thomson opacity and radiation pressure tends to approach a uniform entropy state. 
Assuming hydrostatic equilibrium and an equation of state dominated 
by radiation pressure ($p \propto \rho^{4/3}$),  the density stratification 
in the optically thick envelope can be written as
\begin{equation}
  \rho_{\tau} = \frac{f M_\ast}{4 \pi \ln (r_{\tau}/r_{\rm t}) r^3},
  \label{eq:rho_ext}
\end{equation}
where $f$ is the fraction of  stellar material in the envelope and $r_{\tau}$
is the radius at which the envelope becomes optically thin, which for  electron scattering
opacity is given by
\begin{equation}
  r_{\tau} \approx 1.7 \times 10^{15} \left( \frac{f M_{\ast}}{0.5 M_{\odot}}\right)^{1/2}
  \mathrm{cm.}
  \label{eq:rout}
\end{equation}

The density distribution derived in equation (\ref{eq:rho_ext})  is probably valid only at late times ($t \sim t_{\rm peak}$), as
the returning gas does not immediately produce a thick disk of radiation-dominated gas. First material must enter quasi-circular orbits.
The bound orbits are initially very eccentric and the range of orbital periods is large. The orbital semi-major axis of the most tightly bound debris is 
\begin{equation}
a \approx 10^3 \left({M_{\rm h}\over 10^6 M_\odot}\right)^{-1/3}\left({r_\ast \over R_\odot} \right)\left({M_\ast \over M_\odot}\right)^{-2/3} r_{\rm g},
\end{equation}
and the period is 
\begin{equation}
t_{\rm a} \approx 7.3 \left({a \over 10^3 r_{\rm g}}\right)^{3/2} \left({M_{\rm h}\over 10^6 M_\odot}\right)^{-1/2}\,{ \rm days},
\end{equation}
where $r_{\rm g}\approx 1.5\times10^{11}(M_{\rm h}/10^6 M_\odot)$ cm is the gravitational radius.
As a result of  internal dissipation due to high viscosity or shocks, the debris raining down would, after one or two orbital periods ($t_{\rm a} \ll t_{\rm peak}$), form a highly elliptical disk  surrounded by an extended  atmosphere with a wide spread in apocentric distances, whose progressively increasing vertical scale  is likely to be smaller than the one predicted by equation (\ref{eq:rout}). For a $10^7M_\odot$ black hole, $t_{\rm a}$ at $a\approx r_{\tau}$ is approximately  161 days while $t_{\rm peak}\approx 89$ days.

  \subsection{Medium Shaped by the Interaction of  Stellar Winds}
  \label{sec:win}

Aside from our own galactic center and M31, we have no direct measurement of  the gas content at sub parsec scales near  massive black holes. The stellar density, however, is more well-known. After all, if the stars were not closely packed near the center of the galaxy, we would not have evidence for central black holes within quiescent galaxies. These same stars should provide mass winds, whose strength depends on the concentration of stars enclosed in the black hole's sphere of influence as well as the rates and velocity of the mass injection of those winds. 

To determine the gas structures surrounding quiescent  supermassive black holes, we follow the formalism  developed  by \citet{quataert04}, who modeled  the  distribution of hot gas around  the central parsec of the galactic center under the assumption of spherical symmetry and an adiabatic equation of state. The one-dimensional hydrodynamical equations are also solved using FLASH \citep{fryxell00}, following the method described in \citet{naiman12}. The winds from the closely packed stellar members are assumed to shock and thermalize such that density and energy contributions can be treated as source terms in the hydrodynamical equations. In spherical symmetry, these equations can be written as \citep{holzer70}:
\begin{equation}
\frac{1}{r^2} \frac{d}{dr} \left( \rho_{\rm a} v r^2 \right) = q(r)
\end{equation}
\begin{equation}
\rho_{\rm a} v \frac{dv}{dr} = - \frac{d p}{dr} - \frac{G M_{\rm h} \rho_{\rm a}}{r^2} - q(r) v
\end{equation}
\begin{equation}
\frac{1}{r^2} \frac{d}{dr} \left[ \rho_{\rm a} v r^2 \left(\frac{v^2}{2} + \frac{c_{\rm s}^2}{\gamma-1} 
\right)\right] 
+  \frac{ \rho_{\rm a} v G M_{\rm h}}{r^2} = \frac{q(r) v_{\rm w}^2}{\gamma \left(\gamma - 1 \right)}
\end{equation}
where  the velocity, $v$, and density of the ambient medium, $\rho_{\rm a}$, depend solely on $r$, and 
$c_{\rm s}$ is the sound speed of the gas.

At the central parsec of the galactic nuclei, gas is assumed to be supplied by winds which originate from massive stars. 
In our own galactic center, these stars include blue super giants each with with mass loss rates $\dot{M}_w \approx 10^{-4} M_\odot {\rm yr}^{-1}$ and wind speeds of $v_{\rm w} \approx 600 -1000\;{\rm km\; s^{-1}}$ \citep[e.g.][]{naj1997}.  Here we use the term $q(r)$ to quantify the total rate of mass injection from the stellar winds:
\begin{equation}
\Sigma \dot{M}_{\rm w} = \int 4 \pi r^2 q(r) dr.
\end{equation}

Following \citet{quataert04} we set $q(r) \propto r^{-\eta}$ with  $\eta = 0, 2, 3$ for $r \, \epsilon \, [r_{\rm m}, r_{\rm M}]$ 
and $q(r) = 0$ otherwise.  Different values of $\eta$ correspond here to different mass injection distributions. A value of $\eta=0$, for example, describes mass  that is injected preferentially  at large radii while $\eta=3$ corresponds to equal mass injection for all radii within  $[r_{\rm m}, r_{\rm M}]$. To uniquely specify $q(r)$, we must determine $\Sigma \dot{M}_{\rm w}$ and $\eta$ as well as $[r_{\rm m}, r_{\rm M}]$.   For 
simplicity, we assume the central star cluster properties 
from \citeauthor{quataert04} can be scaled with  the central black hole's mass,  so that 
$[r_{\rm m}, \, r_{\rm M}] \propto M_{\rm h} ^{1/3}$ and $\Sigma \dot{M}_{\rm w} \propto \Sigma N_\ast \propto M_{\rm h}$. 

Figure \ref{fig:mhden} shows the results of our simulations 
for  different black hole masses. After many sound crossing times the flow settles into a steady state. Far away from the cluster, as expected,  the flow is driven out by the aggregate influence of the  stellar winds while interior to the cluster  the gas is captured and accreted onto the central black hole. The stagnation radius, the boundary where the flow is divided between inflowing and outflowing, increases with the mass of the central black hole.  
As can be seen from  Figures \ref{fig:mhden} and \ref{fig:mhdeneta} , the
density stratification is strongly dependent on the black hole
mass but nearly independent on the power-law assumed for the distribution of the
mass injection (i.e. $\eta$). 

The density distribution at small radii can be roughly described  by  $\rho_{\rm a}\propto r^{-k}$ 
with $k\approx 1,1.4,1.5$ for $M_{\rm h}=10^6,10^7,10^8 M_\sun$ respectively. In the outer regions,
the density  is seen to rapidly converge  to a $k=2$  wind profile.
Realistically, we would expect the  ambient medium around the supermassive black hole to have a complex multiphase structure as inferred from models of the line-emitting region \citep{barai11}.

  \section{Propagation of Jets  in Quiescent  Galactic Nuclei}
  \label{sec:jet}

\subsection{The Underlying Dynamics}

A simple analytical argument can be used to understand the 
dynamical evolution of  jets. Let us consider a cylindrically symmetric, relativistic jet moving through
a stratified medium. As long as the flow is continuously injected, the 
head of the jet (or working surface) will have a double shock structure, composed by a 
forward and a reverse shock where, in the system of reference
where the contact discontinuity is at rest, the ambient medium and the jet material respectively
are decelerated and heated, transforming kinetic into thermal energy.

The speed $v_{\rm h}$ with which
the head of the channel advances into the surrounding medium  is obtained
by balancing the momentum flux in the shocked jet material  against that of the
shocked surrounding medium, measured in the frame comoving with the advancing head 
(e.g. \citealt{begelman89, rr02, matzner03, bromberg11, lazzati11, metzger11}):
\begin{equation}
 \rho_{\rm j} h_{\rm j} \Gamma_{\rm j}^2\Gamma_{\rm h}^2\left(v_{\rm j}-v_{\rm h}\right)^2 + p_{\rm j} = 
 \rho_{\rm a} h_{\rm a} \Gamma_{\rm h}^2v_{\rm h}^2 + p_{\rm a} \;,
  \label{eq:vel}
\end{equation}
where $\rho_{\rm j,a}$, $h_{\rm j,a}$, $\Gamma_{\rm j,a}$,  
and $p_{\rm j,a}$ are the mass density, specific enthalpy, Lorentz factor
and pressure of the jet and ambient medium respectively,  and $v_{\rm j,h}$
is the velocity of the jet and the advancing head 
(where $u=\Gamma \beta=\Gamma v/c$  is the proper fluid velocity and $h_{\rm a} \simeq 1$).
If $v_{\rm h}$ is supersonic with respect to the ambient gas, 
$p_{\rm a,j}$ can be neglected and equation (\ref{eq:vel}) gives 
\begin{equation}
\label{eq:beta}
v_{\rm h}={v_{\rm j}}\left[1+\left(\frac{\rho_{\rm a}}{\rho_{\rm j} h_{\rm j}\Gamma_{\rm j}^2}\right)^{1/2}\right]^{-1}.
\end{equation}
The resulting flow  pattern will depend crucially upon the jet Mach
number and the density ratio between the jet and the
given ambient  medium. When the jet
density significantly exceeds the ambient density, $v_{\rm h}\approx v_{\rm j}$.

A relativistic jet emerging from the galactic core  will pass through 
a region of steadily decreasing ambient density  
for up to several tens of parsecs (Figure \ref{fig:diagram}).
Even if the jet is  confined, this decrement in density 
will result in an increase in cross section, and the degree of 
collimation $\theta=s/r$ will either decrease or increase, depending 
on whether the size  of the evacuated channel $s$ increases more or 
less rapidly than $r$. 
This  indicates that the dynamics  of the expanding jet is  expected 
to be  modified by changes in  collimation and can not be properly 
captured  by spherically symmetric solutions. 
These difficulties motivate consideration of the hydrodynamical  
confinement in  axisymmetric numerical calculations, to which we now 
turn our attention.

\begin{figure*}
\centering
\includegraphics[width=\linewidth]{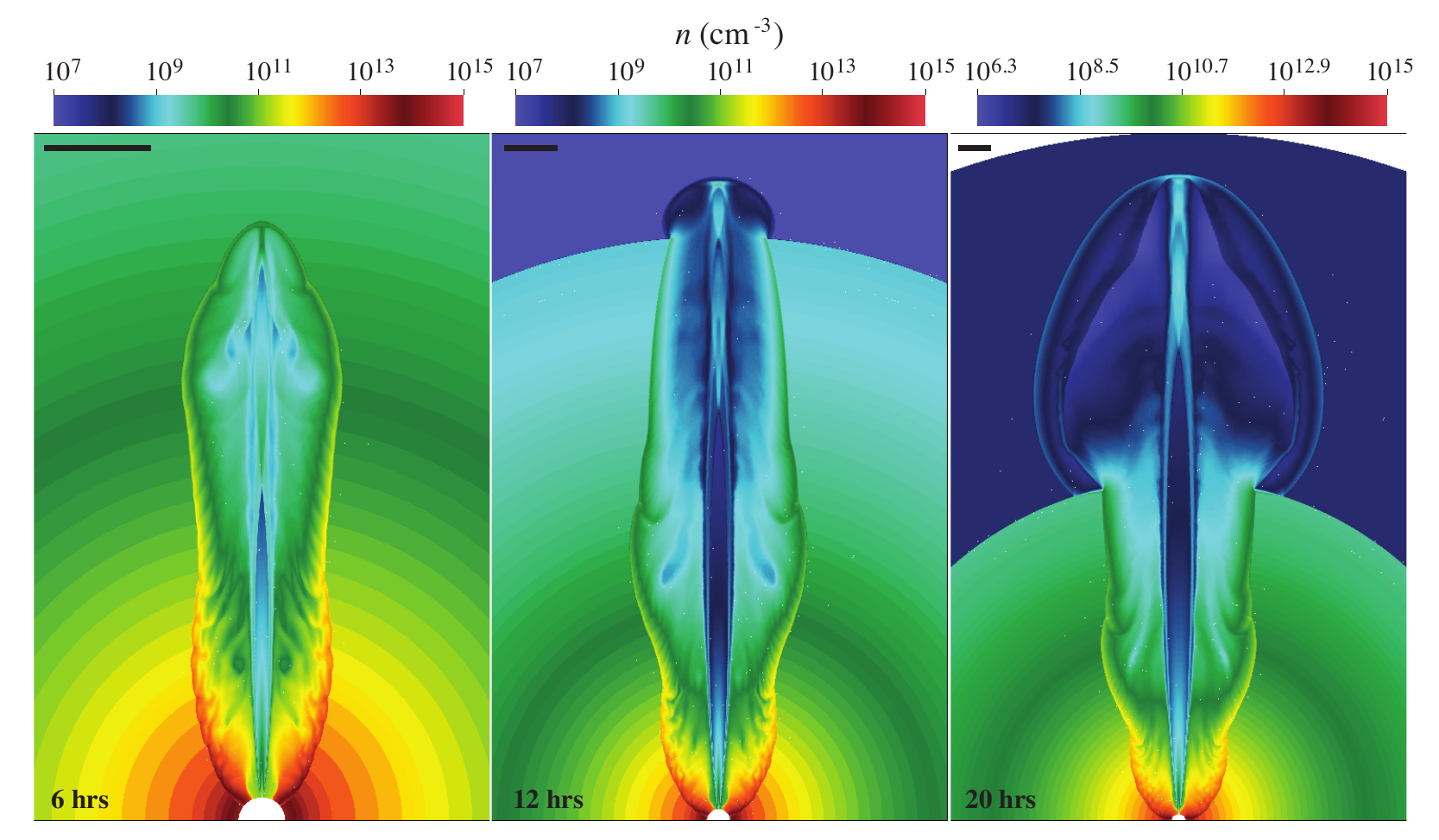}
\caption{Propagation of a relativistic jet through the envelope of high-entropy gas that might form around a massive black hole of mass $M_{\rm h}=10^7M_\odot$ as a result of the tidal disruption of a star. Shown are logarithmic density cuts in cm$^{-3}$. 
Evolutionary ages   after injection  (assumed here to be at $t \sim t_{\rm peak}$) are indicated in each frame
together with a corresponding size $10^{14}$ cm scale bar.  
The jet is initialized at $r_0=r_{\rm t}= 4.6 \times 10^{13}$~cm with $\Gamma=10$ and $\theta_0=0.1$. The  jet  luminosity, characterized by $\epsilon_{\rm j}=0.1$,  is injected as 20\%  thermal and  80\% kinetic. 
The envelope is characterized by $r_{\tau}=  10^{15}$~cm, corresponding  to $f =0.4$. For $r > r_{\tau}$, the density and pressure  are assumed to be constant with
$\rho_{\rm a}= 10^{-2}\rho_{\rm a}(r_{\tau})$ and  $p_{\rm a} = p_{\rm a}(r_{\tau})$. The size of the computational domain was $(2r_{\tau})^{2}$. The simulation employs a two-dimensional spherical adaptive mesh grid, with 
100 $\times$ 40 cells along the $r$ and $\theta$ directions
at the coarsest level, and 5 level of refinement, with
a maximum resolution of $\Delta r = 1.3 \times 10^{12}$~cm 
and $\Delta \theta = 2.5\times 10^{-3}$~rad. }
\label{fig:jet_env}
\end{figure*} 

  \subsection{Numerical Methods and Initial Model}

The propagation of relativistic jets triggered by the tidal disruption 
of stars in quiescent supermassive black holes is studied here by using 
the adaptive mesh refinement code \emph{Mezcal}. The code solves the 
special relativistic hydrodynamics (SRHD) equations in two-dimensional 
spherical (polar) coordinates.  The
{\it Mezcal} code integrates the SRHD equations by using a
second-order upwind scheme. The equation of
state, relating enthalpy to pressure and density,
assumes an adiabatic index $\gamma = 4/3$.
A detailed description of the SRHD version of the \emph{Mezcal} 
code is presented  in \citet{DC11a} together with a 
series of standard tests\footnote{The code is routinely used to calculate 
the dynamics and appearance of  relativistic, impulsive flows, thought to 
accurately describe the evolution of gamma ray burst afterglows \citep{DC11b}.}.

Common to all calculations is the initiation of   a conical jet with an 
initial opening angle $\theta_0$  and uniform velocity $u_0=\Gamma_0\beta_0$,
with matter injected  along the symmetry axis at the inner 
boundary of the computational domain $r=r_0$. The luminosity of the jet is assumed here to follow the mass accretion  rate,  $L_{\rm j} (t) =\epsilon_{\rm j} c^2  \dot{M}_{\rm h} (t)$. 
This relation  together with mass conservation determines the density of the jet, $\rho_{\rm j}(r_0)$, at the injection boundary.  
Given the large range in  scales,  the propagation of the jet in  the optically thick   envelope ($r\leq r_{\tau}$) and its subsequent  
expansion through the  medium shaped by the interaction  of massive stellar winds ($r\leq r_{\rm M}\approx 10^3 r_{\tau}$) are studied separately. 

\begin{figure*}
\centering
\includegraphics[width=\linewidth]{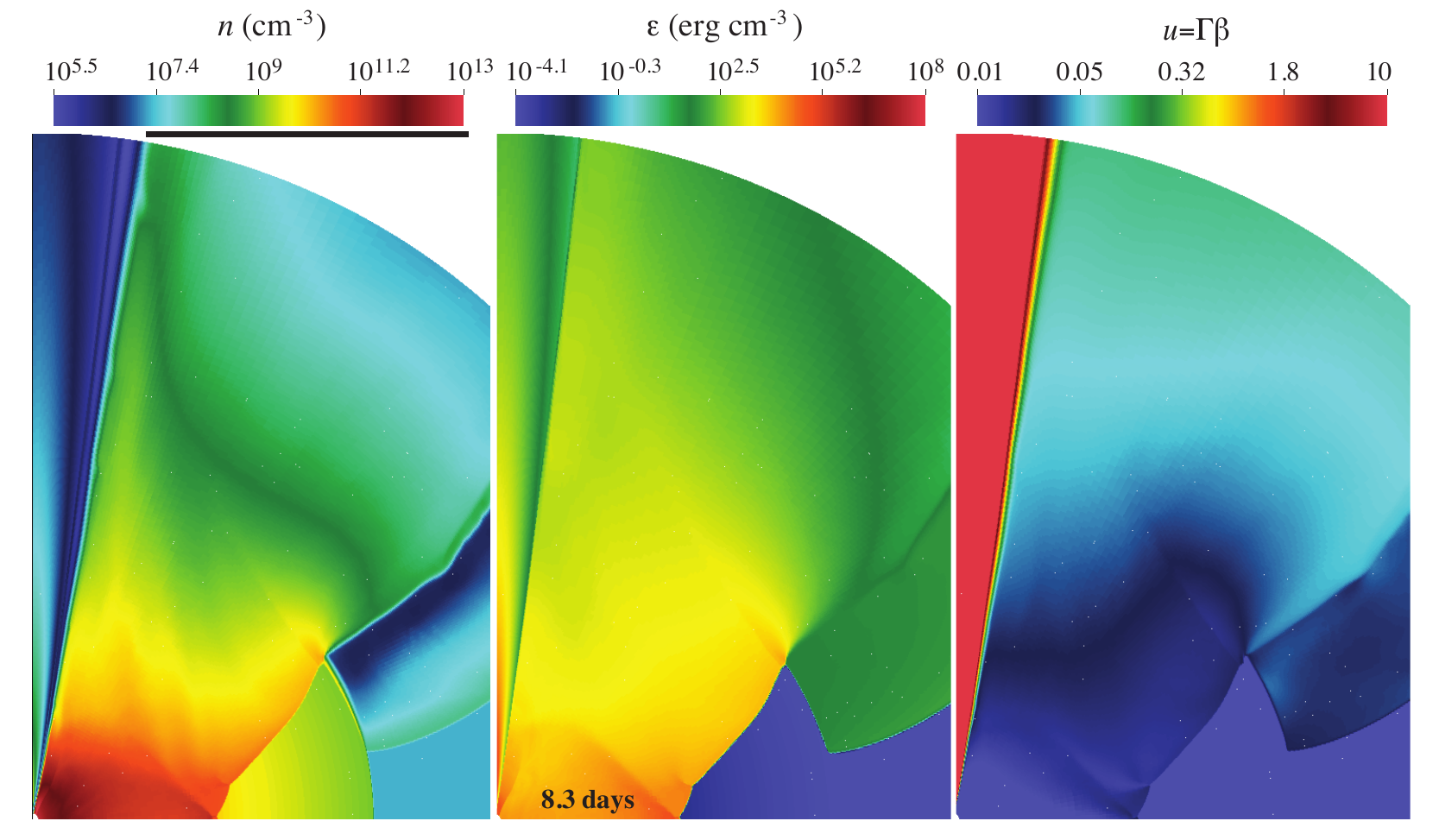}
\caption{The expansion of the cocoon material $\sim 7.9$ days after  the jet head in Figure \ref{fig:jet_coc}  has broken free from the optically thick envelope.  Shown are the
density, internal energy and velocity at 8.3 days after injection (assumed here to be at $t\sim t_{\rm peak}$). A $10^{14}$ cm size scale bar is shown in the left panel.  
}
\label{fig:jet_coc}
\end{figure*} 

  \subsection{Jet Propagation in the  Optically Thick  Envelope}
  \label{sec:jetenv}

The existence of a steady, spherical, optically thick envelope around the black hole should be regarded as an extreme assumption as its  extension and structure will be modulated at a variable rate
by fallback. The envelope starts
to form when the most tightly bound debris fall back. Our stellar disruption simulation (Section \ref{sec:rates}) shows that the first
material returns at a time $t_{\rm a}$, with an infall rate of about $\dot{M}_{\rm a}\approx 10^{-4} M_\odot\;{\rm yr}^{-1}\leq \dot{M}_{\rm Edd}$ (Figure \ref{fig:dmdedens}). The
 infall rates are expected to increase, relatively steadily, for at least tens of orbital periods,
before the majority of the bound material rains down at  $t\approx t_{\rm peak}$. Once the envelope is formed, it will evolve under the influence of viscosity,
radiatively cooling  and time dependent mass injection from both the jetted outflow and the angular momentum redistribution region.

To study the evolution  of the relativistic jet in the optically thick region, we assume, for simplicity, the envelope   to be fully formed ($t\sim t_{\rm peak}$) and accurately described by the steady, spherical hydrostatic solutions presented in Section \ref{sec:env}. 
In addition, since the jet head breaks out  of  the envelope after only a time 
$r_{\tau}/c \ll t_{\rm peak}$, the jet power  is assumed to be steady during this phase with $L_{\rm j} \approx \epsilon_{\rm j} \dot{M}_{\rm peak} c^2$.  

Snapshots of detailed hydrodynamic simulations of the evolution of the jet through the optically thick envelope are presented in Figure \ref{fig:jet_env}, where the density maps of the expanding jet at various times after $t_{\rm peak}$ in its
hydrodynamical evolution are plotted. Initially, the low-density jet is unable to move the envelope material at a speed comparable to its own and thus is abruptly decelerated. Most of the energy output during that period is deposited into a cocoon surrounding the jet, in which the energy supplied by the jet exceeds that imparted to the swept-up envelope material by a factor $\beta_{\rm h}^{-1}$.  In the cocoon region the jet and ambient medium remain separated by a contact discontinuity where shearing instabilities are prominent. 
The cocoon region exhibits two important dynamical effects \citep[e.g.][]{bromberg11}. First, it forms a weak shock that moves laterally  at the relativistic internal sound speed, and second, it acts to partially collimate the jet itself. As the jet expands further into the envelope, the drastic density drop permits the jet head to accelerate to velocities close to the speed of light  ($\beta_{\rm h}\approx 1$). Thus, if it is a general property that the jet becomes relativistic near the boundary of the envelope, the outer edge of the envelope is reached in a crossing time $\approx r_{\tau}/(c\beta_{\rm h}\Gamma_{\rm h}^2$) as measured by an observer along the line of sight.  

\begin{figure*}
\centering
\includegraphics[width=\linewidth]{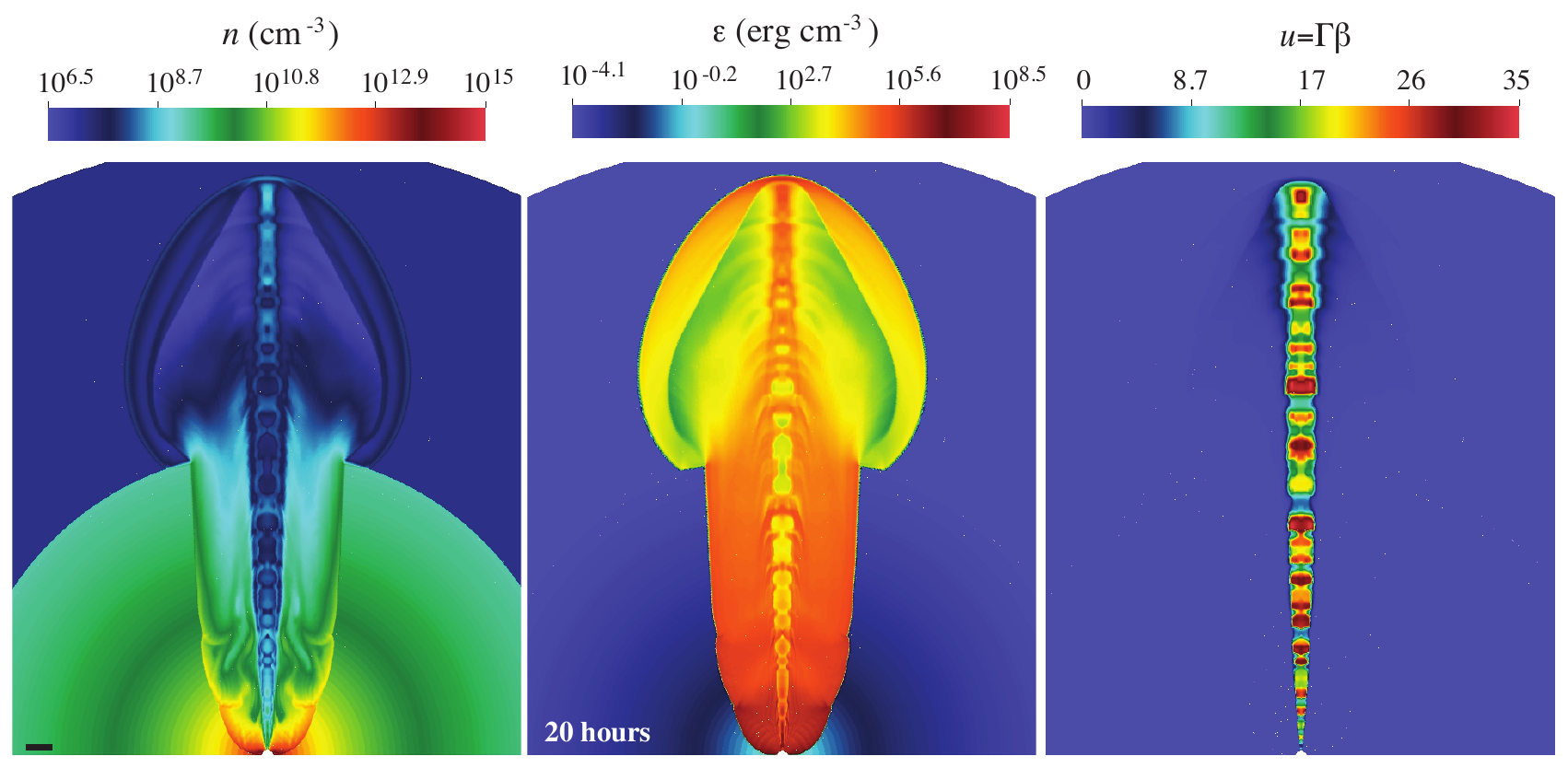}
\caption{Propagation of a relativistic  jet with an imposed random 
variation in both time, $\delta t\;\epsilon\; [10^2,10^3]$ s,  and velocity,  $\Gamma\; \epsilon\; [2,20]$. The jet average properties  are the same as in
Figure  \ref{fig:jet_env}.  Shown are the
density, internal energy and velocity at 20 hours after injection (assumed here to be at $t\sim t_{\rm peak}$).  A $10^{14}$ cm size scale bar is shown in the left panel.  
}
\label{fig:svar}
\end{figure*} 

After the jet has broken free from the envelope region, the fraction of relativistic plasma injected into the cocoon will be much reduced as the jet accelerates and  $\beta_{\rm h} \approx 1$. 
The amount of energy that accumulated in the cocoon while the jet was advancing sub-relativistically is 
\begin{equation}
E_{\rm c}\approx 10^{50}\bar{\beta}^{-1}_{\rm h} \left({r_{\tau} \over 10^{15}{\rm cm}}\right)\left({L_{\rm j}\over 10^{46}{\rm erg\;s^{-1}}}\right){\rm erg},
\end{equation}
where $r_{\tau}/\bar{\beta}_{\rm h}$ is the envelope traversal time and $\bar{\beta}_{\rm h}$ is the average speed of the jet head. The energy accumulated in this phase is thus larger than the binding energy of the envelope\footnote{provided that $M_{\rm h}< 10^8M_\odot$, above which the majority of main sequence stars  are swallowed whole.}. At the radius $\approx r_{\tau}$ where the head of the jet starts to advance relativistically, the volume of the region incorporated into the cocoon is related to both the jet and cocoon expansion velocities by $V_{\rm c}\approx (\pi/3)r_{\tau}^3(\beta_{\rm c}/\beta_{\rm h})^2$. Unlike the jet, this cocoon material does not have a relativistic outward motion, although it has a relativistic internal sound speed. At first an elongated bubble (since pressure balance may never be reached within a radiation-dominated isentropic atmosphere) will be inflated, which can expand most rapidly along the rotation axis and  will eventually unbind the envelope (Figure \ref{fig:jet_coc}). Even under the extreme assumption that the envelope is fully formed at $t\approx t_{\rm peak}$, the energy deposited by the jet is expected to eject most of the envelope material in the optically thick region. Thus, passage through this region cannot significantly  alter the jet launching conditions or enhance collimation.  

  \subsubsection{Internal Shocks}
  \label{sec:is}  
Instead of assuming  the jet to be a steady outflow, here we
suppose that it is irregular on   timescales much shorter than $t_{\rm peak}$. For instance,
if the Lorentz factor in the outflowing collimated ejecta varied by a factor of more
than 2, then the shocks that developed when fast material overtook
slower material would be internally relativistic \citep{rees94}.  Dissipation would
then take place whenever internal shocks developed in the ejecta, which can then be reconverted into energetic particles and radiation (the jet may also lose energy as it propagates through the photon field of the accretion disk). Internal shocks generated either as a consequence of
fluctuations at the source \citep[e.g.][]{kobayashi97,
ramirez-ruiz02} or arising from the development
of large-amplitude instabilities provide an attractive explanation
for the  large scale variability seen in the Swift 1644+57/GRB 110328A event \citep{bloom11, burrows11}.

\begin{figure*}
\centering
\includegraphics[width=\linewidth]{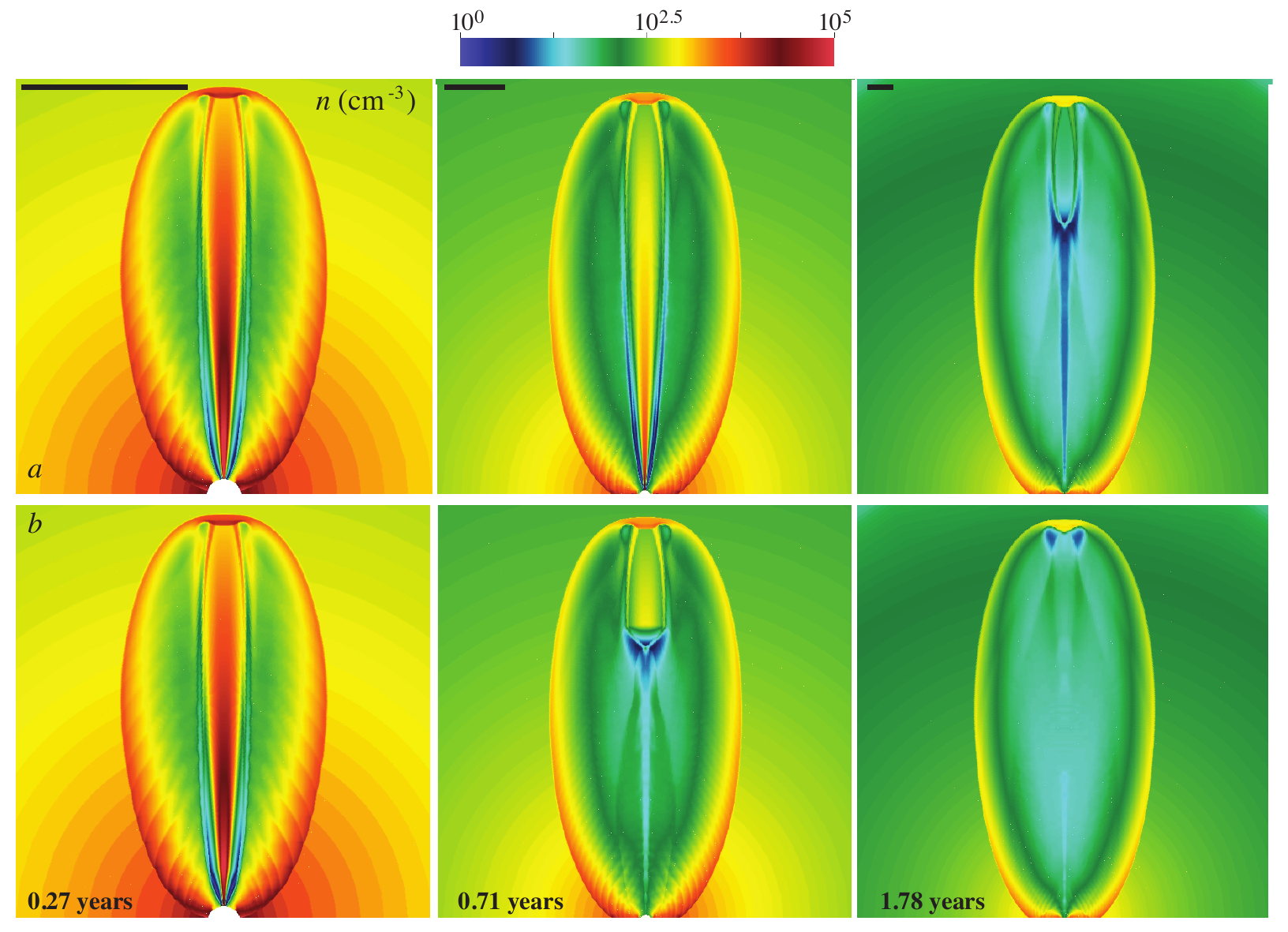}
\caption{
Propagation of a relativistic jet through the sub-parsec region  shaped by the interaction of massive stellar winds that might be present in the nucleus of  a massive black hole of mass $M_{\rm h}=10^7M_\odot$. Shown are
logarithmic density cuts in cm$^{-3}$.  Evolutionary ages   after injection 
are indicated in each frame
together with a corresponding size $10^{17}$ cm scale bar.  
The jet is initialized at $r_0=  10^{16}$~cm with $\Gamma=5$ and $\theta_0=0.1$. The  jet  luminosity, characterized by $\epsilon_{\rm j}=0.1$,  is injected as 20\%  thermal and  80\% kinetic. 
The external density  is  taken from the simulations of the interacting stellar winds shown in Figures \ref{fig:mhden} and \ref{fig:mhdeneta}, while the jet  luminosity is assumed to follow the mass feeding rate  $\dot{M}_{\rm h} (t)$   for as long as  $ \dot{M}_{\rm h} (t)\gtrsim \dot{M}_{\rm Edd}$ (case {\it a}). In case {\it b}, the jet luminosity is  halted when $\dot{M}_{\rm h} = \dot{M}_{\rm peak}$.   Cases {\it a} and   {\it b} have  been constructed to  illustrate  the effect of
the $\propto t^{-5/3}$ injection phase on the jet dynamics. The simulations employ a two-dimensional spherical adaptive mesh grid, with  400 $\times$ 4 cells along the $r$ and $\theta$ directions
at the coarsest level, and with 10 level of refinement, corresponding
to a maximum resolution of $\Delta r = 1.3 \times 10^{13}$~cm 
and $\Delta \theta = 7.7\times 10^{-4}$~rad in the radial and azimuthal 
direction respectively. With this resolution, the azimuthal extent of the jet
is resolved with $\sim 65$ cells at the finest level of refinement.
}
\label{fig:jet-large-1}
\end{figure*} 

 To illustrate the basic idea, suppose that two blobs of equal  mass, but with different
Lorentz factors $\Gamma_i$ and $\Gamma_j$ (with $\Gamma_i >\Gamma_j
\gg 1$) are ejected at times $t_1$ and $t_2$, where $t_2 - t_1 =
\delta t$. In the case of highly relativistic ejecta, the shock
develops after a distance of order 
\begin{equation}
r_{\rm \iota} \approx c \delta t \frac{2 \Gamma_i^2\Gamma_j^2}{\Gamma_i^2-\Gamma_j^2}
\approx 6\times 10^{15} \left({\delta t \over 10^3{\rm s}}
\right) \left( {\Gamma_j\over 10} \right)^2\rm cm.
\end{equation}
The reconversion of bulk energy can
be very efficient: when the two blobs share their
momentum, they move with $\Gamma_{ij}=\sqrt{\Gamma_i\Gamma_j}$, so
the fraction of the energy dissipated is
\begin{equation}
\epsilon_{\iota}={\Gamma_i +\Gamma_j -2\sqrt{\Gamma_i\Gamma_j} \over \Gamma_i
+\Gamma_j}.
\label{eq:effm}
\end{equation}                
High efficiency does not, therefore, require an impact on matter at rest;
all that is needed is that the relative motions in the comoving frame
be relativistic.  

The  evolution of a non-steady  relativistic  jet  with randomly varying Lorentz factor $\Gamma$ (between 2 and 20) over  a range of   timescales  $ 10^2{\rm s} \lesssim \delta t  \lesssim 10^3$s is shown in Figure~\ref{fig:svar}. The average parameters of the jet  are  chosen so that they are similar  to those displayed in Figure~\ref{fig:jet_env}. Compared to Figure~\ref{fig:jet_env}, significant structural differences along the evacuated channel appear when faster material catches up with  slower material   and a strong shock forms. An internal  shock in the relativistic jet will move with a Lorentz factor of up to $\Gamma_{ij}\approx 10$, and the emitting material behind the shock will be subject to a large Doppler boost. Dissipation, to be most effective, must occur when the envelope  is optically thin, i.e. $r_\iota \gtrsim  r_\tau$ (Figure~\ref{fig:svar}).  

The deceleration of the working surface allows slower ejecta to catch up with the head of the jet, replenishing and reenergizing the
reverse shock and boosting the momentum in the working surface. Since the efficiency of converting bulk motion to radiation in an internal shock depends on the difference 
between $\Gamma$'s, the fact that the working surface is decelerating ($\beta_{\rm h}\lesssim 1$) implies that the efficiency
will be much higher than in the standard case where $\Gamma$ is assumed to fluctuate by a factor of a few within the jet and among blobs on a  typical timescale
 $\delta t \gtrsim r_{\rm g}/c \approx 50 (M_{\rm h}/10^7 M_\odot)$ s.

  \subsection{Jet Propagation in the Stellar Wind Region}
  \label{sec:jetwind}

We now turn our attention to the  evolution of the jet as it expands 
through the large scale  environment of the galactic nuclei  where  the density stratification is thought to be regulated by the interaction of winds from the surrounding massive stars. The deceleration  of the  jet within this region gives rise to the non-thermal, long-lived radio emission observed  in the Swift 1644+57/GRB 110328A event \citep{zauderer11,berger11}.  The density profile within this region is 
taken from the simulations of the interacting stellar winds described in Section \ref{sec:win}
and shown in Figures \ref{fig:mhden} and \ref{fig:mhdeneta}. The power of the jet is assumed to follow the mass feeding rate,  $L_{\rm j} (t) =\epsilon_{\rm j} c^2  \dot{M}_{\rm h} (t)$, with $ \dot{M}_{\rm h} (t)$ taken from the results of the tidal disruption simulations presented  in Section \ref{sec:rates} and Figure \ref{fig:dmdedens}.  

\begin{figure}[t]
\centering
\includegraphics[width=\linewidth]{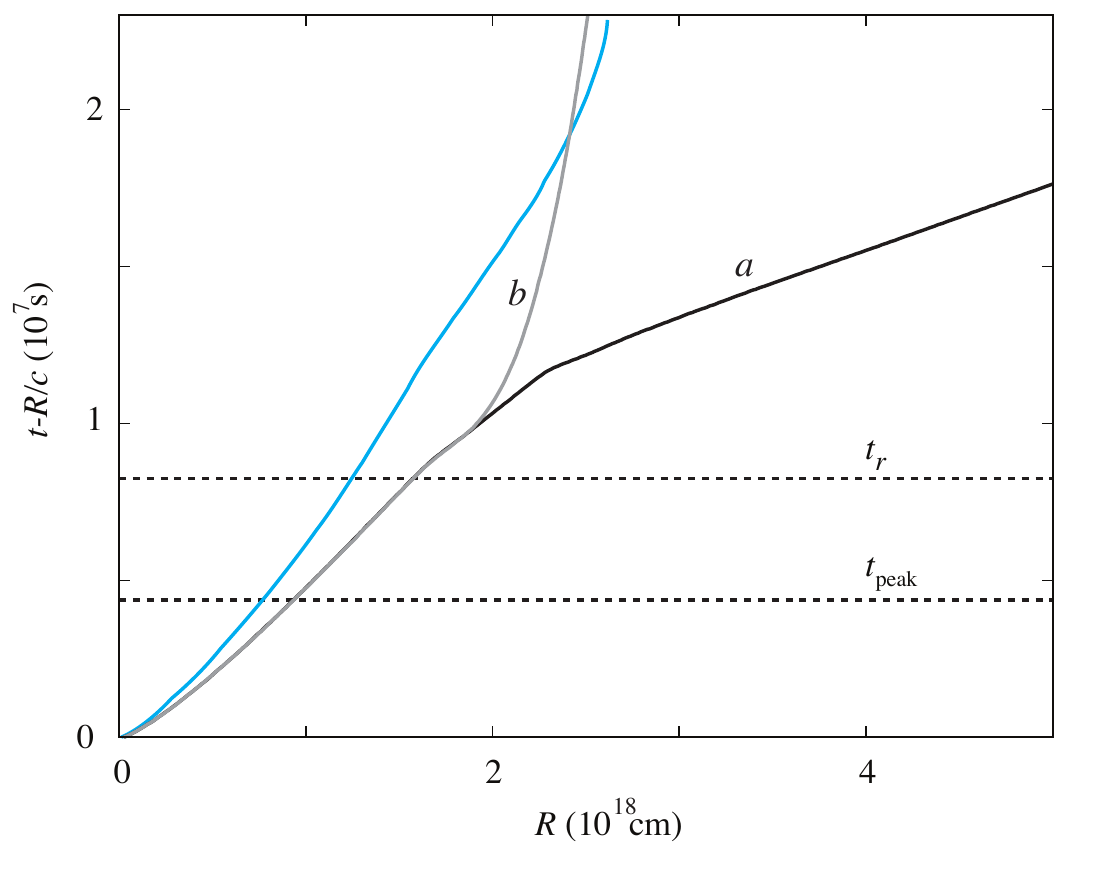}
\caption{Schematic space time diagram in source frame coordinates for the two cases depicted in Figure \ref{fig:jet-large-1}. The axes (logarithmic) are {\it r} versus $t-(r/c)$, where $t$ is time measured in the source frame, and is zero when the
star is initially  disrupted. In this plot, light rays are horizontal lines. Shown are the
position of the bow shock as a function of time.
In case {\it a},  $L_{\rm j} \propto \dot{M}_{\rm h} (t)$  for $t\leq t_{\rm Edd}$. In case {\it b}, the jet luminosity is  halted  at $t=t_{\rm peak}$. Information about the abrupt change in the feeding rate (at $t\gtrsim t_{\rm peak}$)  propagate
into the jet at the internal sound speed and are communicated to the jet by an inward propagating 
rarefaction wave that reaches the jet head at $t_{\rm r} \gg t_{\rm peak}$. 
The blue solid line shows, for comparison, the evolution of the working surface as predicted by equation (\ref{eq:vel}) for case {\it b} under the assumption of an non-evolving jet opening angle.}
\label{fig:jet-large-2}
\end{figure} 

Little is known about the relation between jet production in the more massive accreting black holes and the state of their constituent accretion disks. However, a similar association of hot thick accretion flows with the strongest jets as seen  in binary black holes is  expected.  Motivated by this, we assume that strong relativistic jets preceding tidal disruption are  only triggered  when  $ \dot{M}_{\rm h} (t)\gtrsim \dot{M}_{\rm Edd}$.  The properties of these jets are thus severely constrained by the mass feeding rate history. The emitted radiation is an observable diagnostic which provides constraints on the  processes occurring at their point of origin.

The propagation of a  jet  injected at the onset of the super-Eddington accretion phase over successive decades in radius ranging from $10^{-2}$ pc  to  1 pc  is shown in 
Figure \ref{fig:jet-large-1} for $M_{\rm h}=10^7M_\odot$. Two illustrative cases  are depicted.  In case {\it a}, the jet power  increases with time to achieve a maximum
at $t=t_{\rm peak} \approx 0.27$~yrs, it then  subsequently decreases as $\sim t^{-5/3}$ until it is  finally halted at the time when $\dot{M}_{\rm h} (t) < \dot{M}_{\rm Edd} \approx 0.71$~yrs. In case {\it b}, the injection history is the same until the jet power is prematurely halted  at  $t=t_{\rm peak}$.  Figure~\ref{fig:jet-large-2} shows the corresponding  schematic world-lines of the bow shock evolution for the two illustrative cases depicted in Figure~\ref{fig:jet-large-1}.

As seen in  Figure \ref{fig:jet-large-1}, the jet  evacuates a channel out to some location where it impinges on the surrounding medium.  A continuous flow of relativistic fluid emanating from the nucleus supplies this region with mass, momentum and energy. The jet, unable to move the surrounding material at a speed comparable to its own, 
is decelerated and, as a result,  most of the energy output during that period is deposited into a cocoon surrounding the jet. As a first approximation, evolution of the cocoon's
shape is governed both by the advance of the head and by
the cocoon's own pressure-driven sideways expansion into
the surrounding medium.  Heavy jets therefore propagate almost ballistically and
are naturally surrounded by small cocoons. Light jets
have large cocoons and, as seen in Figure \ref{fig:jet-large-1}, 
can  be confined by the pressure of the shocked material. 

The cocoon  will  effectively collimate  the jet \citep{bromberg11} as long as
\begin{equation} 
 \rho_{\rm j} h_{\rm j} \Gamma_{\rm j}^2/\rho_{\rm a} < \theta_0^{-4/3} \;.
 \label{eq:coll}
\end{equation}
where  $\rho_{\rm j} h_{\rm j} \Gamma_{\rm j}^2/\rho_{\rm a}$ gives the ratio between the jet's energy density  and the rest-mass energy density of the surrounding medium at the location of the head.  Collimation is then seen to increase with decreasing $k$ for $k\leq 2$. When  $k=2$  the jet's head velocity is constant  and consequently most of the energy flowing into the cocoon  can not effectively   counterbalance the jet's expansion.
The density profile $ \rho_{\rm a} \propto r^{-k}$ in the stellar wind mass injection region  (region $II$ in Figure~\ref{fig:diagram})
varies  from $1.4 \lesssim k \lesssim  2$ with increasing $r$ for a $M_{\rm h} = 10^7M_\odot$ and, as a result,
the cocoon's pressure is expected to collimate the jet and decrease its opening angle. This is clearly illustrated  in  Figure~\ref{fig:jet-large-2} in which the results of the simulations are directly compared with  the evolution of the working surface as predicted by equation (\ref{eq:beta}). In a jetted  source
like  Swift 1644+57/GRB 110328A, the expansion of the working surface  within the  gas on scales $r\leq 1$ pc is clearly incompatible with that predicted on the assumption of a constant  jet opening angle.  

\begin{figure}
\centering
\includegraphics[width=\linewidth]{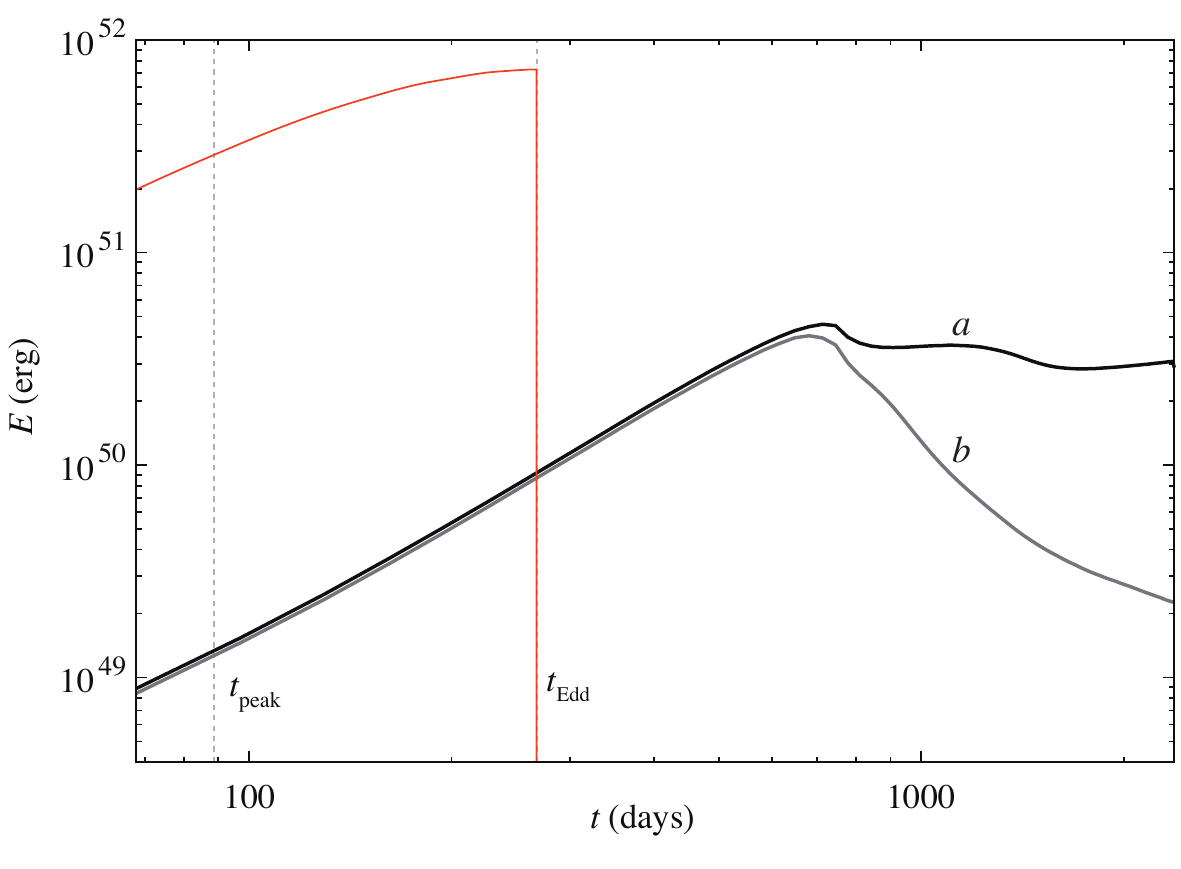}
\caption{The evolution of the  total internal energy available at the working surface for  the two cases depicted in Figure \ref{fig:jet-large-1}.  Also shown (red line)  is the cumulative energy injected by the relativistic jet as a function of time.
After rarefaction wave reaches the jet head, which occurs at similar times in both cases,  the available  internal energy drops although it does so more abruptly  in case {\it b} where the jet power  is halted at  $t_{\rm peak}< t_{\rm Edd}$.   }
\label{fig:jet-large-3}
\end{figure} 

If the jet were suddenly to turn off (in case {\it a} this occurs  at $t=t_{\rm Edd}$ while  in case {\it b} at $t=t_{\rm peak}$) it would be preceded by the collapse of the jet channel which, suddenly evacuated, would be filled in on the transverse sound crossing time by whatever material happened to surround it. This ambient or
cocoon material could be driven into the jet channel at  the internal sound speed (Figure~\ref{fig:jet-large-1}). After the evacuated channel collapses, a cylindrical
rarefaction wave will propagate along the jet channel, reaching the head 
at $t=t_{\rm r}\gg t_{\rm peak}$ when the jet has traveled a much larger  distance ($r_{\rm r}$)  than when it was  initially turned off. This is
followed by a series of compression and rarefaction waves
of decreasing amplitude.  In case {\it a}, the inclusion of the $\propto t^{-5/3}$ injection phase  produces  only   a small change in   $t_{\rm r}$. This is clearly illustrated in Figure~\ref{fig:jet-large-3}, where we have plotted  the total internal energy available at the working surface. After the rarefaction wave reaches the jet head, which is observed to take place at comparable times in both of the cases depicted in Figure~\ref{fig:jet-large-1},  the available  internal energy drops  although it does so less  abruptly in case {\it a} where the jet continues to be powered  until $t_{\rm Edd}$. The intensity of the radiation from the radio-emitting electrons  which have recently emerged from the hot spot  rapidly decreases  after a time  $t_{\rm r}\gg t_{\rm peak}$, which corresponds to a time $\approx r_{\rm r}/(c\beta_{\rm h}\Gamma_{\rm h}^2)$ as measured by an observer along the line of sight. 

\begin{figure}
\centering
\includegraphics[width=\linewidth]{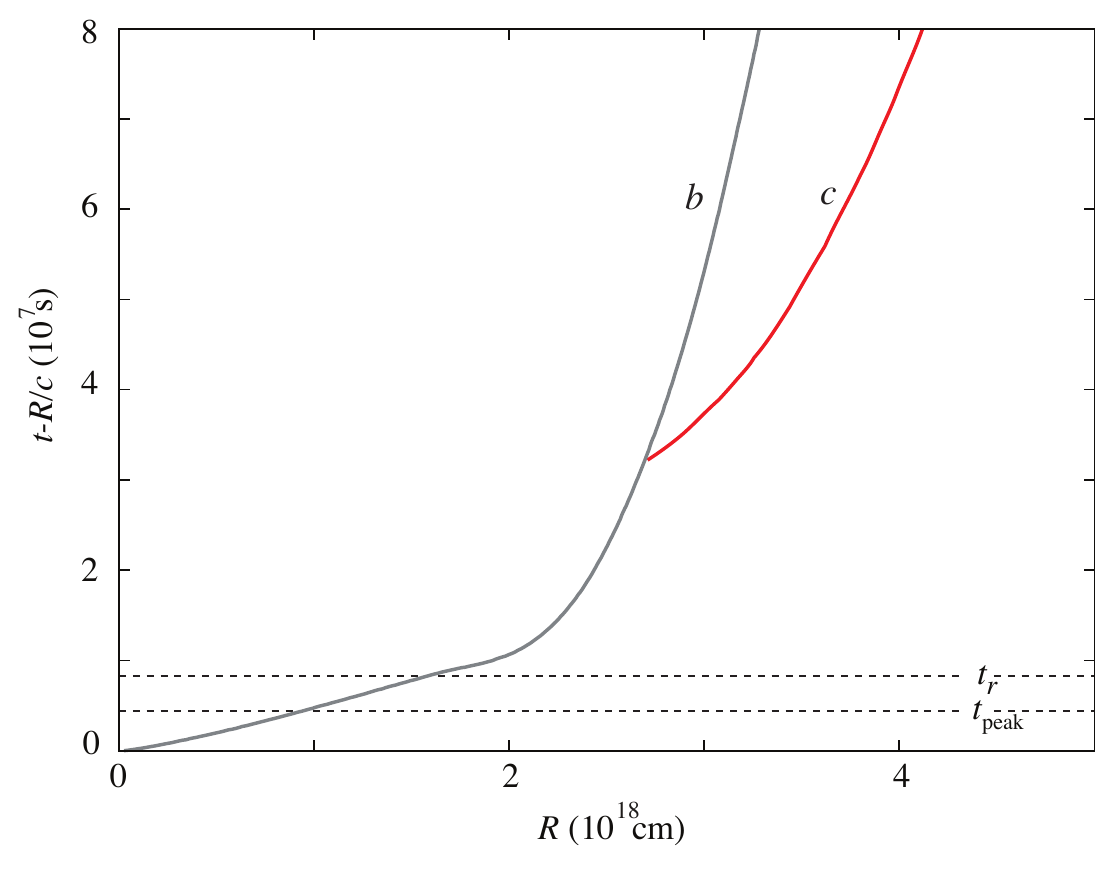}
\caption{Schematic space time diagram in source frame coordinates illustrating the position of the bow shock.  The simulation is the same
as in Figure \ref{fig:jet-large-2}  for case {\it b}  (jet injection stopped at $t=t_{\rm peak}$) 
but with significant energy injection arising  from slowly moving material (case {\it c}).
}
\label{fig:jet-large-4}
\end{figure}

When the jet becomes free at $t \gtrsim t_{\rm r}$, the necessary condition for the jet  to remain in pressure balance with its surroundings is no longer  satisfied and strong shocks are driven into the jet. As a result, the contact discontinuity and the forward shock are  abruptly decelerated. This deceleration allows ejecta to catch up and pass through a reverse shock just inside the contact discontinuity,  increasing the dissipated internal energy at the shock front (Figure~\ref{fig:jet-large-4}) and in principle  giving rise to a longer-lived afterglow than that predicted by the communication delay between the working surface  and the base of the jet. 

  \section{Discussion} 
  \label{sec:discussion}

\subsection{The  Properties of Jets Triggered by  Tidal Disruption}
\label{sec:summary}
Quiescent black holes appear to be capable of producing  directed flows of relativistic matter.  Given the twin requirements of  super Eddington luminosity and short flaring timescales, the currently favored models involve  the  disruption of a star with the subsequently accreted debris providing the required luminosity.   Here we examine, within this model,  the consequences of assuming a relation between jet production and the state of their constituent accretion disks. In those sources that display jets, production of the jet should only occur when the disk is in a hard accretion state and  not
when the disk emission is soft. Testing this
prediction in quiescent galactic nuclei, as argued in Section \ref{sec:demographics}, will require statistical studies  because the typical jet production  duty cycle time is expected to be thousands of years. Confirmation of this effect will provide additional evidence that accretion states similar to those in binary black holes also should exist in supermassive black holes.  
As we show in Section \ref{sec:swift}, many of the observed properties the Swift 1644+57/GRB 110328A event can be understood  as resulting from a hot accretion flow, driven into a  previously quiescent supermassive  black hole, which is then subsequently collimated into a pair of anti-parallel jets. 
If we were to venture a general classification scheme for tidal disruption jets, we would obviously expect the black hole mass, the rate at which the gas is supplied to the black hole, the angular momentum of the black hole, the flow velocity and the orientation relative to our line of sight to be the essential parameters. 

Most of the radiation we receive is  reprocessed by matter quite distant from the black hole.
The fact that jets are detectable at all means that some re-randomization of kinetic energy and re-acceleration of particles must be occurring along their length to counteract radiation and adiabatic losses.
Under the assumption that these prime movers are able to form collimated, relativistic outflows, the observable effects of the two major radiating regions, the working surface and internal shocks, are studied. This leads to a unified picture where internal shocks (and/or external Compton at the base of the jet)  can provide most of the variable, high photon energy luminosity, and where 
the region of greatest radio emissivity is  located at the head of the jet. 

Hot radio spots are naturally interpreted as the working surface at the end of a jet. The jet is decelerated at a strong collisionless shock where particle acceleration and field amplification can occur. The relativistic electrons and magnetic field pressure are balanced by the ram pressure of the ambient medium. Because the source moves relativistically, aberration of light must be accounted for when calculating the observed radio light curve as as well as the local direction of polarization. And as the radio hot spot is unlikely to be resolved, we can only measure the average polarization over the whole  image. As a result, the breaking of symmetry of the emitting region around our line of sight  is required  in order to produce a net polarization.  Statistical studies over a sample of  tidal disruption jets, or time resolved polarimetry of different emission episodes within a single event, may teach us about the dominant emission mechanism, the jet structure, or the magnetic field configuration within the hotspot.

It is obvious from the discussions in this paper that the dynamics of
tidal disruption jets  are complex, especially because of rich interactions
between the jet and the surrounding medium.  Confirmation was provided of the important notion
that the visibility of jets, in particular the emission emanating from the head of the jet, is  determined largely by the luminosity and velocity injection history  of the transient  jet as well as the properties of the  surrounding medium.
Axisymmetric hydrodynamical calculations of these jets show that the resulting
dynamics are different from those predicted by one-dimensional models; interaction 
with the surrounding medium can not only enhance collimation but also mediate the communication delay between the base  and  the head of the jet. 
Even in the simplest case of a jet  whose thrust directly scales 
with $\dot{M_{\rm h}}$, complex behavior with multiple possible transitions in the observable 
part of the radio afterglow lifetime may be seen. The eventual resulting afterglow light curve depends 
fairly strongly on the properties of  $\dot{M_{\rm h}}$, especially the impact parameter $\beta$ and mass 
of the star and  the black hole. There is a good and bad side to this. On the negative side, it 
implies that one can not be too specific about the times at which we expect to see transitions in 
the observed emission. On the positive side, if and when we do see these transitions, they can be 
fairly constraining on the properties of the system.

\subsection{On the Nature of Swift 1644+57/GRB 110328A}\label{sec:swift}

The X-ray and radio flux from the Swift 1644+57/GRB 110328 event has been interpreted as being emitted by the jet produced from the super-Eddington accretion of material resulting from the disruption of a star by a supermassive black hole. For the first $\sim 30$ days since the flare triggered the BAT instrument, the object exhibited tremendous variability on $\lesssim 10^{4}$ s timescales, with isotropic equivalent flare luminosities $> 10^{48}$ ergs/s in the soft X-rays alone. Beyond this first month of violent activity, the degree of variability decreased, and appeared to follow the predicted $t^{-5/3}$ power law associated with the late-time evolution of the fallback \citep{berger11}.

The initial month-long light curve plateau, punctuated by periods of extreme variability, has been presumed to be associated with the phase of the event in which the accretion rate exceeds the Eddington limit \citep{berger11,metzger11}. However, as the Eddington limit is $\sim 100$ times smaller than the peak accretion rate for a complete disruption by a $10^{6} M_\odot$ black hole, only a disruption by a more massive black hole or a partial disruption, for which the peak accretion rates are comparable to the Eddington rate, are capable of producing such a short period of super-Eddington accretion (Figures \ref{fig:dmdedens} and \ref{fig:mdottime}). But while a partial disruption or a disruption by a heavier black hole can yield a plateau of the appropriate duration, these events peak at a much later time relative to the time of disruption, resulting in a decay slope that yields a poor match to the observed X-ray light curve.

\begin{figure}
\centering
\includegraphics[width=\linewidth]{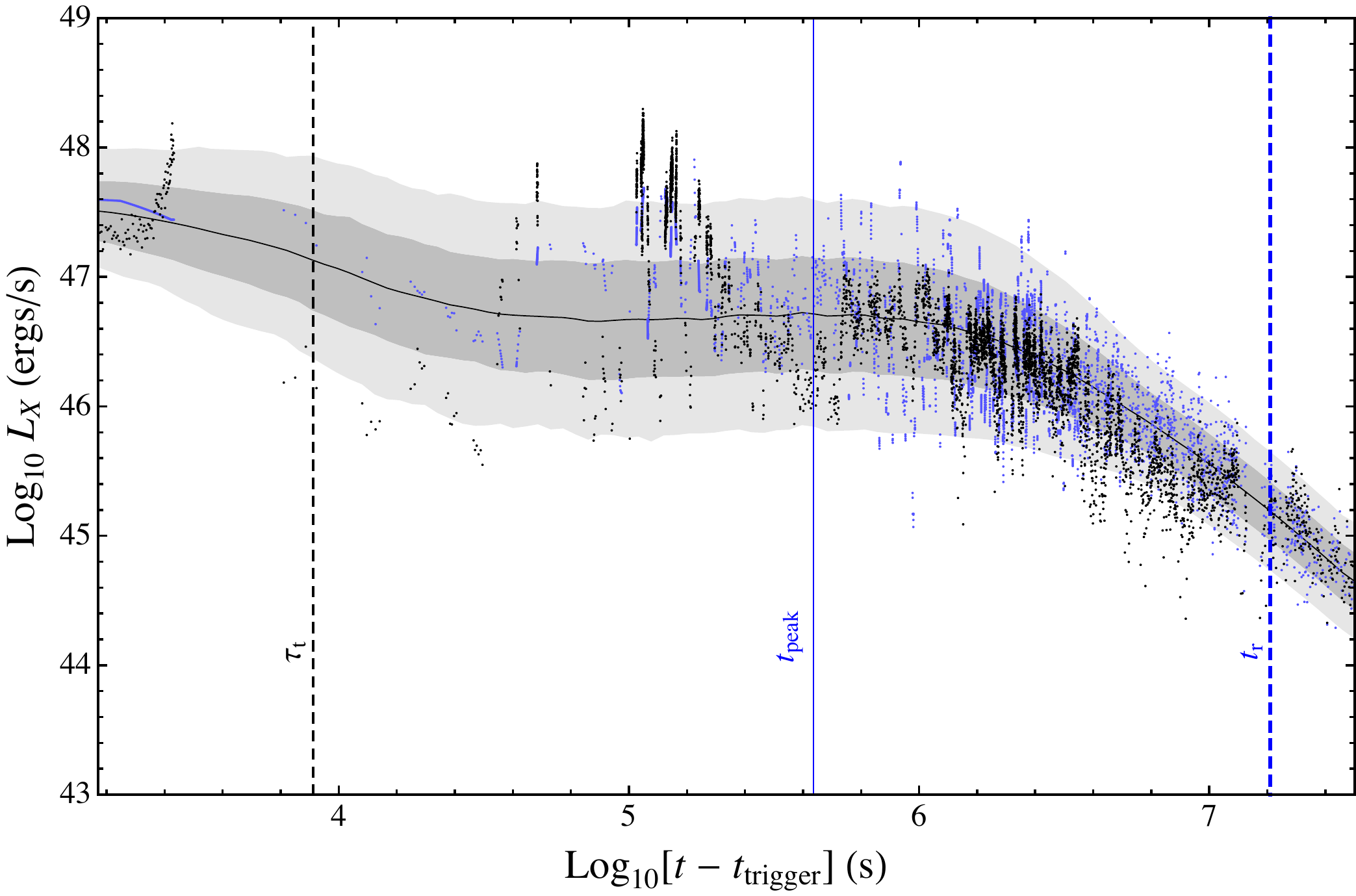}
\caption{Luminosity in X-rays $L_{\rm X}$ from 2 - 10 keV for Swift J1644+57. The black points show the measured flux from the event, whereas the blue points show the realization from our stochastic model of jet emission from $10^{3}$ realizations of a flare generated by the disruption of a $1 M_{\odot}$ star by a $10^{6} M_{\odot}$ black hole that bears the most resemblance (i.e., the minimum $\chi^{2}$) to the observed light curve. The sample times for the chosen realization are selected to be equal to the observation times of the Swift event. The gray contours show the 1- and 2-$\sigma$ variances from the mean luminosity $L_{\rm j}$, which is shown as the solid black curve. The black dashed vertical line shows $\tau_{\rm t}$, the orbital period at the tidal radius, the characteristic timescale of variation in our model. The blue vertical lines show timescales associated with the presented realization, with the thin line showing $t_{\rm peak}$, whereas the thick line shows \(t_{\rm r}\) as calculated by equation (\ref{eq:beta}).}
\label{fig:swift}
\end{figure}

A proposed solution to this problem is to allow the star to penetrate more deeply at pericenter \citep{cannizzo11}, for which analytical approximations in which the binding energy distribution across the star is presumed to be ``frozen in'' at pericenter \citep{evans89,lodato09} can produce shorter flare times. These shorter flares could reproduce both the month-long plateau and the observed decay rate. However, numerical simulations by \cite{guillochon12} have shown that an increase of $\beta$ beyond the value for which the star is completely destroyed do not result in faster transients, with the time of peak actually tending to slightly {\it larger} values with increasing $\beta$ for complete disruptions. For very deep encounters in which general relativistic effects are important, \cite{laguna93} show that while $t_{\rm peak}$ tends to smaller values, it does not scale as sharply as predicted by the energy freezing model, and the time at which a $\beta = 5$ and a $\beta = 10$ encounter cross the Eddington limit is almost identical. Therefore, a deep encounter is incapable of reproducing the duration of the two phases exhibited by the X-ray light curve.

We suggest an alternative model in which the flare is a standard full disruption of a main sequence star, but that the triggering of the BAT instrument and the month-long plateau are actually artifacts of variability driven at a timescale comparable to the orbital period at the tidal radius, $\tau_{\rm t} = 2\pi\sqrt{r_{\rm t}/GM_{\rm h}}$. In this model, $\dot{M}_{\rm h}$ exceeds Eddington   even during the observed $t^{-5/3}$ decay phase. Variability on this timescale could be driven by the interaction of the material that returns to pericenter with previously accreted material within the accretion disk. Hydrodynamical studies of the fallback stream have shown that the material that returns to pericenter after a disruption passes through a small nozzle region where it is compressed violently \citep{rosswog09,ramirez-ruiz09}, resulting in a rapid heating of material that may affect the nozzle dynamics or the dynamics of the accretion disk. As both the production of the jet and the seed photons emitted by the disk are dependent on the instantaneous value of $\dot{M}_{\rm h}$, the variability introduced by the fallback can have non-linear feedback effects which may be able to produce the observed variations in flux, as is often seen in three-dimensional magnetohydrodynamical simulations of thick accretion disks \citep[e.g.][]{hawley2009}

\begin{figure*}
\centering
\includegraphics[width=0.475\linewidth]{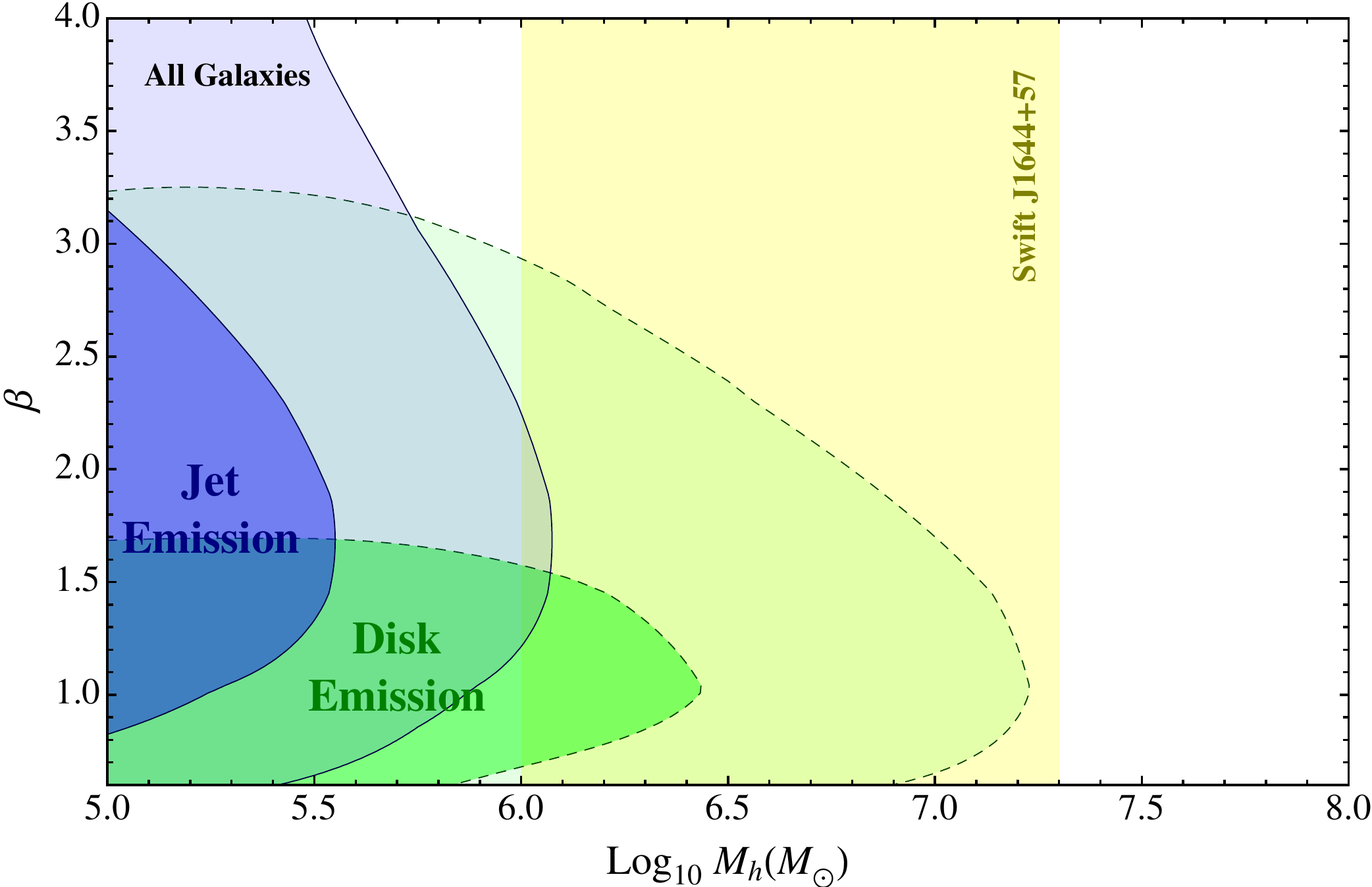}\includegraphics[width=0.475\linewidth]{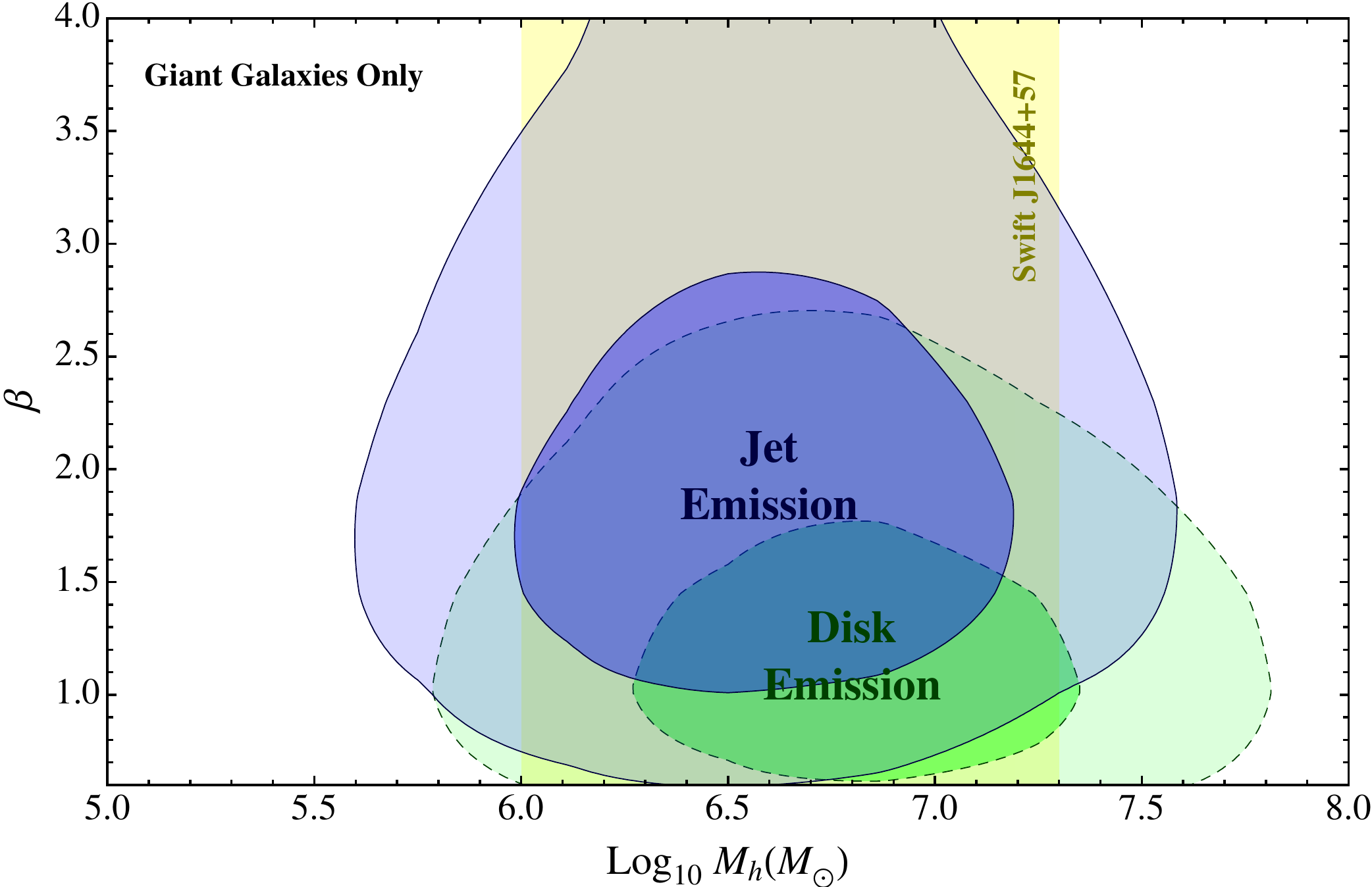}
\caption{Fraction of photons received from the disk and jet emission associated with the tidal disruption of a 1 $M_{\odot}$ stars as a function of the black hole mass $M_{\rm h}$ and impact parameter $\beta$. The green and blue regions show the likelihood contours for the disk and jet respectively, with the dark and light contours corresponding to 50\% and 90\% of all photons received. The disk contribution is calculated as in \citet{strubbe09}, whereas the jet emission is assumed to scale as \(\epsilon c^{2} (\dot{M} - \dot{M}_{\rm Edd})\). The left panel shows the probability contours considering the contribution of black holes residing in both giant and dwarf galaxies, as characterized by Trentham and Tully 2002, assuming a black hole mass to bulge relation of $M_{\rm h} = 10^{-2.91} M_{\rm bulge}$ \citep{merritt01}. The right panel shows the probability contours for giant galaxies only. As $\dot{M} \propto M_{\rm h}^{-1/2}$ and smaller black holes can foster deeper encounters where $r_{\rm p} \ll r_{\rm t}$, a large majority of jet-produced photons are emitted by disruptions occurring around black holes of mass $10^{7} M_{\odot}$ or less, even when the contribution from dwarf galaxies is completely excluded. The limits on $M_{\rm h}$ for the black hole thought to be responsible for Swift J1644+57 are indicated by the yellow region \citep{burrows11}.}
\label{fig:jet_disk}
\end{figure*}

To model this variability, we use the method of \cite{kelly11}, in which the stochastic luminosity variation is modeled as a single Ornstein-Uhlenbeck process with characteristic frequency $\omega = 2\pi/\tau_{\rm t}$, which is then added to the mean luminosity $L_{\rm j}\propto \dot{M}_{\rm h}$. To match the basic character of the observed light curve, we adjust the amplitude of the driving noise depending on the degree to which $L_{\rm j}$ exceeds the Eddington ratio, with a minimum noise amplitude chosen to match the observed variability at late times. At late times, the driving noise we have selected is within a factor of a few of what is measured for steady-state AGN.

Figure \ref{fig:swift} shows the results of this simple model in comparison to the Swift event for $10^{3}$ random realizations  of a flare generated by the disruption of a $1 M_{\odot}$ star by a $10^{6} M_{\odot}$ black hole, along with the realization that happens to give  the best overall fit.  The observed power, assumed to follow $\dot{M}_{\rm h}(t)$, is amplified by two powers of the Doppler factor, $\delta$, with $\delta^2=20$  and with 10\% of the luminosity assumed to be emitted in the 2 -- 10 keV energy range. For each realization, the triggering time is set when the isotropic equivalent luminosity of $10^{47.5}$ erg/s is surpassed. In general, this occurs close to the peak luminosity for the choice of parameters we have selected, with the trigger itself being caused by the apex of one of the mini-flares associated with our variable component. This leads to an envelope shape that shows a subtle decline from the initial triggering event within a few $\tau_{\rm t}$, and then flatten outs over tens of days until $t >$ a few times $t_{\rm peak}$. If the light curve near $t_{\rm peak}$ is characterized by a large degree of variability, as our models predicts, the triggering event occurs either before or after $t_{\rm peak}$ with almost equal probability, and thus identifying the actual time of peak within the light curve itself would be difficult, if not impossible. 

As argued in Section  \ref{sec:jetwind}, even for the simplest assumption of $L_{\rm j}\propto \dot{M}_{\rm h}$, we found that the resulting structure and dynamics of the radio hot spots are very different from those predicted by the standard spherical solutions \citep{bloom11,metzger11,berger11}. This is mainly because at early stages when $t\leq t_{\rm peak}$ the  jet evolution  is governed both by the advance of the head and by the cocoon's  pressure-driven sideways expansion into the ambient medium. The rapid decrease in luminosity at $t_{\rm peak}$ would be preceded by the collapse of the jet channel. The collapse of both the channel and the cocoon would proceed from the galactic nuclei outward. A model in which the hot spots are simply turned off  is  incompatible with  hydrodynamical collapse of the source.
The fact that information about the rapid decrease in luminosity at $t \gtrsim t_{\rm peak}$  is communicated  to the jet head at $t_{\rm r} \gg t_{\rm peak}$ (Figure \ref{fig:swift}) provides an attractive explanation (without the need of late time energy ejection) for the  delay  between the jet power x-ray  luminosity  and the radio afterglow (thought to be produced by the synchrotron emitting electrons at the jet's head) seen in  the Swift 1644+57/GRB 110328A event \citep{berger11}. The most difficult task at present is to predict  $t_{\rm r}$, the time  at which the intensity of the radiation from the radio-emitting electrons  which have recently emerged from the hot spot  starts to rapidly decrease, which corresponds to a time $\approx r_{\rm r}/(c\beta_{\rm h}\Gamma_{\rm h}^2)$ as measured by an observer along the line of sight. 
 This is mainly because the hydrodynamical collapse of the channel depends sensitively  on the initial Lorentz factor and opening angle of the jet as well as on  the poorly known state of the surrounding environment, from the initial density gradient created by stellar wind injection  to the large-scale ambient  structures transversed by the jet's head at $r\gtrsim r_{\rm r}$.

\begin{figure*}
\centering
\includegraphics[width=\linewidth]{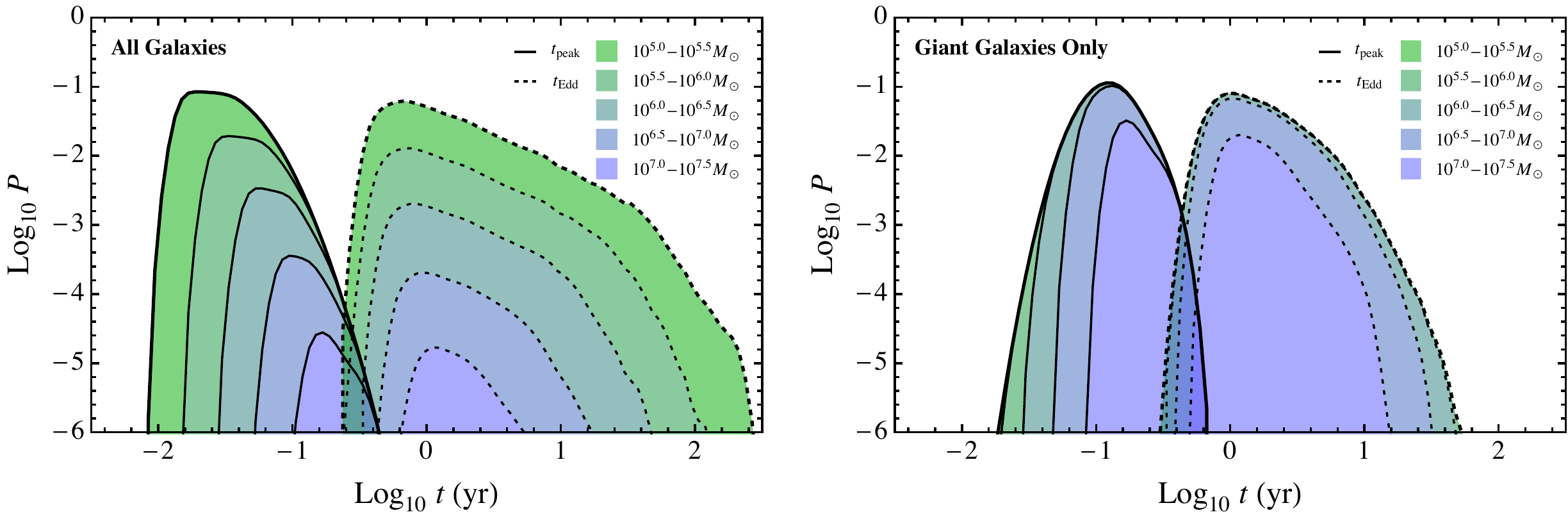}
\caption{Stacked probability histograms showing contribution to measured peak times $t_{\rm peak}$ and time at which the accretion rate falls below the Eddington limit $t_{\rm Edd}$. Each colored region corresponds to a range of black hole masses as annotated in the upper right of each panel, with the solid lines corresponding to $t_{\rm peak}$ and the dashed lines corresponding to $t_{\rm Edd}$. The left and right panels show the same two galaxy distributions as described in the caption for Figure \ref{fig:jet_disk}.}
\label{fig:timehist}
\end{figure*}

  \subsection{Rates, Lifetimes and  BH Demographics}
  \label{sec:demographics}

The rate of transient events coming as the result of the tidal disruption of a main 
sequence star is highly dependent on the population of black holes and their immediate 
environments. The steady-state rate of stars entering the loss cone has been estimated 
by many authors \citep{bahcall76, lightman77, shapiro77, syer99, wang04}, but these 
calculated rates ignore mass segregation, the mass function of stars within the sphere 
of influence \citep{bahcall77}, the importance of black hole mergers \citep{chen09, stone11} ,
and interactions between the central cluster and the host galaxy 
\citep{magorrian99, merritt09}, all of which may affect the rate of stellar 
disruption. As implied by the number of AGN flares thought to result from tidal 
disruptions \citep{gezari09}, the average rate of disruption is known to be 
$\sim 10^{-5}\;{\rm events}\;{\rm yr}^{-1}\;{\rm gal}^{-1}$, but how these events are 
distributed among galaxies remains an open question. Given these uncertainties, we assume 
that the rate of disruption is not dependent on either black hole or galaxy properties 
for the remainder of this section.

Jets are thought to form in optically thick accretion flows resulting from accretion 
rates that are in excess of Eddington \citep{quataert01, narayan08}. The mass function 
of stars around black holes may be substantially different than the standard Kroupa 
IMF function \citep{kroupa01}, but the average stellar mass is on the order of 
$1 M_{\odot}$ for all black holes. As a result, the peak accretion rate simply scales 
with the orbital period at the tidal radius, which low-mass black holes can more easily 
produce jets from the disruption of a star given their smaller Eddington limits. 
This biases the production of jets to low-mass black holes, of which our census is 
growing \citep{ramya11, xiao11} but is still 
too small to definitively relate to galactic velocity dispersion as has been demonstrated 
convincingly with black holes of $M_{\rm h} \gtrsim 10^{6} M_{\odot}$ \citep{merritt01,gultekin09,beifiori11}. 
Emission resulting from disks, on the other hand, is thought to extend 
well below the Eddington rate, and thus the only requirement to produce a transient disk 
from the disruption of a star is that the pericenter of the orbit does not lie within 
the black hole's Schwarzschild radius. As a result, disks are not likely to be produced 
by disruptions for black holes of mass $\gtrsim 10^{8} M_{\odot}$.

Figure \ref{fig:jet_disk} shows the fraction of the total bolometric flux received by an observer from 
both the jet and disk components resulting from the tidal disruptions of a $1 M_{\odot}$ 
stars. The majority of stars that are fed into the black hole originate near the black 
hole's sphere of influence from the {\it pinhole} regime of disruption \citep{lightman77}, 
and thus the differential contribution of stars as a function of the impact parameter 
$\beta$ approximately scales as $\beta^{-2}$. The left panel shows the expected contribution 
of events to each emission type assuming that black holes of all masses are equally common, 
an assumption which is not unreasonable given the potentially flat black hole mass function 
at low black hole mass as measured by \citet{greene07}. If the trend of 
$M_{\rm h}-\sigma$ continued to lower black hole masses, this plot would be further biased 
to black holes of lower mass given the large population of low-mass galaxies. If the trend 
does not continue and black holes of $\lesssim 10^{6} M_{\odot}$ are rare, both emission 
components are biased to black holes of larger mass, as shown in the right panel of 
Figure \ref{fig:jet_disk}. The limits on the mass of the black hole associated with the 
transient event Swift J1644+57 coming from the variability timescale and the $M_{\rm h}-\sigma$ 
relation \citep{bloom11, burrows11} agree quite well with a black hole mass function that 
excludes low-mass black holes, but given the associated uncertainties in the mass determination even a completely flat distribution of black hole masses 
cannot be completely ruled out.

An additional constraint on the progenitors of events similar to Swift J1644+57 is the time at which $\dot{M}_{\rm h}$ reaches its peak value, $t_{\rm peak}$, and the time at which $\dot{M}_{\rm h}$ drops below $M_{\rm Edd}$, $t_{\rm Edd}$. Under the same assumptions used to generate Figure \ref{fig:jet_disk}, we show in Figure \ref{fig:timehist} the probability $P$ of observing a flare with a given value of $t_{\rm peak}$ and $t_{\rm Edd}$. We find that the most probable $t_{\rm peak}$ is a few to tens of days, whereas the most probable $t_{\rm Edd} \sim 1$ -- 10 years, depending on whether or not  low mass central black holes are common. Both distributions are consistent with the Swift event, which our model suggests was caught within days of the peak, and remained above the Eddington limit for $\sim 1$ year.\\

The discovery of flaring black hole candidates in nearby galaxies by Swift will continue to elucidate the demography of the quasar and AGN population, while also enabling us to distinguish between various theoretical models of tidal disruption. As the feeding rate evolves dramatically over only a period of weeks to months, the powering of these formerly dead quasars offers a unique perspective into how the production of jets from black holes depends on the rate at which they are fed and the state of the environment in which they reside. These events can provide valuable insight into the physical mechanisms that operate near massive black holes that are not continually active, a prospect that is otherwise only possible through the study of the nearest few massive black holes. As the number of events increases, the range of possible models that can simultaneously explain the luminosity, variability, color, and time-evolution of these events will diminish, enabling a firm characterization of one of the dominant feeding mechanisms of massive black holes in the local universe.

  \acknowledgments 

We have benefited from many useful discussions with E. Berger, J. Bloom, D. Giannos, J. Grindlay, D. Kasen, B. Kelly, W. Lee, B. Metzger, S. Piran, A. Socrates, A. Soderberg, J. Trump, and A. Zauderer. This research was supported by the David and Lucille
Packard Foundation (ERR, JG and FDC), the NASA Earth and Space Science Fellowship (JG), and the NSF (ERR and JN) (AST-
0847563).


\end{document}